\documentclass[preprint2]{aastex}

\usepackage{natbib} 
\usepackage{graphicx}

\slugcomment{ApJ in press}
\shorttitle{Detailed AGN XLF}
\shortauthors{Miyaji et al.}

\begin{document}
\title{Detailed Shape and Evolutionary Behavior of the X-ray Luminosity Function of Active Galactic Nuclei}   
\author{T. Miyaji\altaffilmark{1,2},
           G. Hasinger\altaffilmark{3},          
           M. Salvato\altaffilmark{4},         
           M. Brusa\altaffilmark{4,5,6},
           N. Cappelluti\altaffilmark{6}, 
           F. Civano\altaffilmark{7,8}, 
           S. Puccetti\altaffilmark{9},
           M. Elvis\altaffilmark{8},                                                 
           H. Brunner\altaffilmark{4},
           S. Fotopoulou \altaffilmark{10},          
           Y. Ueda\altaffilmark{11},                    
           R. E. Griffiths\altaffilmark{12},
           A. M. Koekemoer\altaffilmark{13},
           M. Akiyama\altaffilmark{14}, 
           A. Comastri\altaffilmark{6},
           R. Gilli\altaffilmark{6},
           G. Lanzuisi\altaffilmark{6},
           A. Merloni\altaffilmark{4},
           C. Vignali\altaffilmark{5,6}
}
   \altaffiltext{1}{Instituto de Astronom\'ia sede Ensenada, Universidad Nacional Aut\'onoma de M\'exico,
     Km. 103, Carret. Tijunana-Ensenada, Ensenada, BC, Mexico 
     (mailing address: PO BOX 439027, San Ysidro, CA 92143-9026, USA)}
   \altaffiltext{2}{University of California San Diego, Center for Astrophysics and Space Sciences,
             9500 Gilman Drive, La Jolla, CA 92093-0424, USA}
   \altaffiltext{3}{Institute for Astronomy, 2680 Woodlawn Drive, University of Hawaii, Honolulu, HI 96822, USA
              \email{hasinger@ifa.hawaii.edu}}
   \altaffiltext{4}{Max-Planck-Institut f\"ur extraterrestrische Physik, Giessenbachstrasse 1, D--85748 
     Garching bei M\"unchen, Germany}
   \altaffiltext{5}{DIFA - Dipartimento di Fisica e Astronomia, Universita' di Bologna, viale Berto Pichat 6/2 40127 Bologna, Italy}
   \altaffiltext{6}{INAF --  Osservatorio Astronomico di Bologna, via Ranzani 1, I--40127 Bologna, Italy}
   \altaffiltext{7}{Yale Center for Astronomy and Astrophysics, 260 Whitney Avenue, New Haven, CT 06520, USA} 
   \altaffiltext{8}{Smithsonian Astrophysical Observatory, 60 Garden Street, Cambridge, MA 02138, USA}
   \altaffiltext{9}{ASI Science Data Center, via Galileo Galilei, 00044, Frascati, Italy}
   \altaffiltext{10}{University of Geneva, chemin d’ Ecogia 16, 1290 Versoix, Switzerland}   
   \altaffiltext{11}{Department of Astronomy, Faculty of Science, Kyoto University, Kitashirakawa-Oiwake-cho,Sakyo-ku, Kyoto 606-8502, Japan}
   \altaffiltext{12}{Physics \& Astronomy, STB-216, U. Hawaii at Hilo, Hilo, HI 96720}
   \altaffiltext{13}{Space Telescope Science Institute, 3700 San Martin Drive, Baltimore, MD 21218, USA}
   \altaffiltext{14}{Astronomical Institute, Tohoku University, 6-3 Aramaki, Aoba-ku, Sendai 980-8578, Japan}
%
\email{miyaji@astrosen.unam.mx}

\begin{abstract}  
 We construct the rest-frame 2--10 keV intrinsic X-ray luminosity function of Active Galactic Nuclei (AGNs) from 
a combination of X-ray surveys from the all-sky {\sl Swift} BAT survey to the Chandra Deep Field-South. We use 
$\sim 3200$ AGNs in our analysis, which covers six orders of magnitude in flux. The inclusion of {\sl XMM} and {\sl Chandra} COSMOS data 
has allowed us to investigate the detailed behavior of the XLF and evolution. In deriving our XLF, we take into account realistic AGN spectrum templates, absorption corrections, and probability density distributions in photometric redshift. We present 
an analytical expression for the overall behavior of the XLF in terms of the luminosity-dependent density evolution, smoothed two 
power-law expressions in 11 redshift shells, three-segment power-law expression of the number density evolution 
in four luminosity classes, and binned XLF.  We observe a sudden flattening of the low luminosity end slope of the 
XLF slope at $z\ga 0.6$. Detailed structures of the AGN downsizing have been also revealed, where the number density curves 
have two clear breaks at all luminosity classes above $\log L_{\rm X}>43$. 
{The two break structure is suggestive of two-phase AGN evolution, consisting of major merger  
triggering and secular processes.}
\end{abstract}



\section{Introduction}

 The AGN/QSO luminosity function and its evolution with cosmic time are key observational quantities for understanding
the origin of supermassive black holes (SMBHs) and accretion onto them. This is one of the most important
observational product along with the demography of SMBHs, which are made by scaling relations with host galaxy properties 
(e.g. \citealp{magorrian98,haering04,gultekin09}; see \citealp{kormendiho13} for recent review) as well as  
emerging AGN clustering measurements, which locate typical masses of dark matter halos (DMHs) that the AGNs reside 
\citep[e.g.,][]{mullis04,yang06,coil09,gilli09,cappelluti10,krumpe10,krumpe12} or the halo occupation distribution 
\citep[e.g.,][]{miyaji11,allevato12,richardson13,chatterjee13}.

 X--ray surveys are practically the most efficient means of finding active galactic nuclei (AGNs) over a wide 
range of luminosity and redshift. X-ray emission from almost all extragalactic point X-ray sources with luminosities of 
$L_{\rm X}\ga 10^{42}{\rm erg\,s^{-1}}$ are considered to originate from SMBH accretion, because X-ray emission from 
the star-formation origin (through supernova remnants and X-ray binaries) of galaxies with the highest star formation rates 
typically has a luminosity lower than this \citep{persic04,ranalli05,ptak07}. Enormous efforts have been made by several groups to 
follow up the survey X--ray sources with major optical telescopes around the globe, so that we have fairly complete 
samples of X--ray selected AGNs \citep[][for review]{brandt_has05}.  One of the most important products of such X-ray 
AGN surveys is the X-ray luminosity function (XLF) and its evolution with cosmic time. Since X-ray emission marks the accretion activity onto the SMBHs, the XLF and its evolution gives a bird's eye view of the moments of SMBH growth process. 

 The launch of {\sl ROSAT} and the enormous optical followup efforts of the X-ray sources detected by this satellite 
enabled us, for the first time, to probe the X-ray emitting AGN populations to highly cosmological distances. From the
mid-1990's to mid-2000's, the evolution of the XLF in the soft band (0.5--2 keV) was investigated by numerous 
authors \citep[e.g.,][]{boyle93,page97,jones97,miyaji00a,miyaji01,hasinger05}. While the soft-band surveys found 
predominantly unobscured (type 1) AGNs except at very high redshifts, imaging surveys at  $2<E [{\rm keV}]\la 10$ with 
{\sl ASCA}, {\sl XMM-Newton} and {\sl Chandra} enabled us to expanded the investigation to include obscured (type 2) AGNs, 
predominantly  within the Compton-thin limit ($\log N_{\rm H}\;{\rm [cm^{-2}]}\lesssim 24$), providing a view of the SMBH accretion 
process with much better sampling. Analysis of the $2<E [{\rm keV}]\la 10$ XLF has been made by various groups using 
different samples with different approaches and different levels of sophistication 
\citep[e.g.,][]{ueda03,cowie03,lafranca05,silverman08,ebrero09,yencho09,aird10,ueda14}. Many groups took a simple approach 
and constructed the 2--10 keV XLF assuming a simple power-law spectrum without absorption corrections. From the combination of 
AGNs from {\sl HEAO-1}, {\sl ASCA} and early {\sl Chandra} survey samples, 
\citet{ueda03} constructed an absorption-corrected  2-10 keV XLF with a full $N_{\rm H}$ function, as well as a 
new AGN population synthesis model of the X-ray background. This work was recently revised by \citet{ueda14} 
(herefter, U14), which included both soft and hard samples from diverse sources derived from {\sl ROSAT}, {\sl MAXI}, 
{\sl ASCA}, {\sl XMM-Newton} and {\sl Chandra} surveys. U14 included sophisticated spectral model templates, a revised  
population synthesis model, discussions on possible contributions from Compton-thick AGNs as well as the growth of SMBHs. 
A few works have included absorption into analysis \citep{lafranca05,ebrero09}. \citet{aird10} took a simpler approach in 
spectral assumptions, but used sophisticated statistical techniques and took into account the probability density 
distribution function (PDF) of photometric redshifts and incompleteness corrections, with Bayesian and Markov Chain 
Monte-Carlo methods in their model parameter search. In an attempt to make the XLF construction as free from absorption 
corrections as possible, S. Fotopoulou et al. (2015, submitted to A\&A) restricted their analysis to the  5--10 keV band and use the full Bayesian 
approach in obtaining parameter constraints.      

 Most authors fitted their XLFs to simple analytical formulae. As a common feature of most such models,
the XLF at a given redshift has a (smoothed) two power-law form with a  shallower slope at the low luminosity end
and a steeper slope at the high luminosity end. This is quite unlike the Schechter function, which is customarily used to
describe the optical luminosity function of galaxies and has an exponential decline at higher luminosities. As for the 
XLF evolution, early evolution models with the pure luminosity evolution scheme \citep[PLE; e.g.,][]{boyle93,page97} have been 
superseded by more complex expressions. Many authors have used the Luminosity-Dependent Density Evolution (LDDE) 
\citep[e.g.,][]{schmidt83,miyaji00a,ueda03,hasinger05} model of one form or another. \citet{yencho09} and \citet{aird10} 
considered the models where a smoothed two power-law XLF evolves both in the luminosity and density directions,  
keeping the XLF shape the same. These are  called the (Independent) Luminosity And Density Evolution 
(called the ILDE or LADE) models. Because the change of at least the low luminosity slope with redshift is observed to be 
present, generally the LDDE model is preferred, as critically compared by S. Fotopoulou et al. (2015, submitted to A\&A).  

  When AGN number densities are viewed as a function of redshift in different luminosity classes, 
the AGNs evolve in an ``anti-hierarchical'' manner, or show ``AGN downsizing'' \citep[e.g.,][]{ueda03,cowie03,hasinger05}, 
i.e., the peak of the AGN number density appears at lower redshifts for low luminosity AGNs than those with high 
luminosities. 
This was in contrast to an early theoretical prediction based on AGN triggering by hierarchical merging 
\citep{wyithe03}, in which it was predicted that lower luminosity AGNs peak at higher redshifts (or emerge earlier
in the cosmic history). In recent years, there have been growing support for the hypothesis that the AGN population 
might be composed of two populations that have been triggered by different mechanisms. One is a major merger driven  
triggering, which mainly accounts for high luminosity QSOs and dominant in high redshifts. The other is a secular process, 
which might include fly-by encounters, minor mergers, and/or disk instabilities, and is dominant in intermediate-low 
luminosity \citet[e.g.][]{draper12,georgakakis14}. A number of theoretical works, involving cosmological N-body and/or 
N-body+hydrodynamical simulations and semi-analytical treatments in identifying AGNs, have reproduced and/or explained this 
AGN down-sizing effect \citep[e.g.,][]{dimatteo03,hopkins05,croton09,marulli08,degraf10,fanidakis12}. 

Some models also include ``radio mode'' (a.k.a. ``hot halo'' mode). \citet{fanidakis12} put emphasis on dividing 
the AGN activity into ``starburst mode'', which includes SMBH accretion following both merger-driven and disk instability 
driven starbursts, and the ``hot halo'' mode. By also including empirical obscuration effect, they reproduced the AGN down 
sizing effect. With this general scheme,  \citet{fanidakis13} explained the tendency that intermediate X-ray 
luminosity AGNs are associated with more massive dark matter halos (DMHs) than luminous QSOs 
\citep{miyaji07,krumpe10,allevato11,krumpe12}.  In order to critically compare these lines of theories
and observation, refinements in calculations of the XLF and investigating its detailed evolutionary behavior continues to
be an important observational task.  

 In this work, we use AGNs selected at $E>2$ keV from a collection of surveys. 
Our aim is to construct and investigate the detailed behavior of the XLF. In this particular work,
we make our best effort to construct the luminosity function of unabsorbed and absorbed 
AGNs within the Compton-thin limit, i.e. absorbing column densities of 
$\log N_{\rm H}{[\rm cm^{-2}]}\lesssim 24$.
 Above this column density, Compton scattering causes photons to travel along longer paths within the absorbing 
medium and are subject to much higher chance of photo-electric absorption. In such Compton-thick (CTK) AGNs, 
X-ray emission at $E\lesssim 10$ keV, which is observable with  {\sl ASCA}, {\sl XMM-Newton}, and {\sl Chandra}, is 
highly suppressed unless they are at very high redshifts. Thus AGNs detected in the $E\lesssim 10$ keV surveys are 
considered to be dominated by Compton-thin (CTN) AGNs, with the exception of a small number of X-ray sources, for which 
X-ray emission is dominated by in scattered and reflected components \citep{brunner08,brightman12,brightman14}. 
In the local universe,  higher energy X-ray surveys  ($20\la E\la 200$ keV) such as those available from 
{\sl Swift} BAT \citep[e.g.,][]{tueller10,ajello12}, {\sl INTEGRAL} \citep[e.g.,][]{beckmann09,treister09} can 
detect modestly Compton-thick (CTK) AGNs $24\la \log N_{\rm H}[{\rm cm^{-2}}]\la 25$ and quantify their 
space density and emissivities. {\sl NuSTAR} \citep[e.g.,][]{alexander13} is extending this to higher redshifts ($z\sim 0.5-1$).
  
  We present here a new estimate of the X-ray luminosity function evaluated at the traditional 2--10 keV 
rest frame. Complementary to a recent work by \citet{ueda14}, 
whose emphasis is on the X-ray population synthesis of the X-ray background as well as discussions 
on the role of the CTK AGNs, our emphasis is on the detailed behavior of the shape and evolution of the XLF as well 
as discussions of possible systematic errors.  In particular, in order to assess the uncertainties due to 
photometric redshifts, we include the PDFs of photometric redshifts (photo-z's)
into analysis and compare with the results obtained by using single best-fit photo-z's.  We focus on the X-ray 
luminosity function of unabsorbed and Compton-thin (CTN) AGNs because of the wide availability of  
$E\lesssim 10$ [keV] AGNs.  Also X-ray luminosity $L_{\rm X}$, which is the independent
variable of the XLF presented in this paper, is primarily defined as the {\it intrinsic} 
2-10 keV luminosity, i.e., before absorption and not including the reflection component.
 
 In this work, the sample is greatly expanded by the addition of the datasets from {\sl XMM}-COSMOS 
\citep{hasinger07} and C-COSMOS \citep{elvis09}, which are parts of the COSMic evolution Survey 
(COSMOS)\citep{scoville07}. In Sect. \ref{sec:cosmos}, we explain our COSMOS sample, including the construction  
of our combination of the {\sl XMM}-COSMOS and C-COSMOS datasets.  The details of the selection criteria for other 
samples from the literature are explained in Sect. \ref{sec:global_sample}. At the end of this section, we explain 
our incompleteness correction and derive the $N(>S)$ function for extragalactic X-ray sources in
each sample. 

 In sect. \ref{sec:xlf_calc}, we explain the computation of our estimated XLFs. Model parameter
estimations using a Maximum-likelihood fitting method, which also include absorption and the probability 
density distributions (pz-PDFs) of photometric redshifts, are explained. This section also covers 
the $N_{\rm obs}/N_{\rm mdl}$ estimation of the binned XLF in the presence of the absorption and pz-PDFs. 

 Sect. \ref{sec:xlf_global} presents our global expression of the Luminosity-Dependent Density Evolution (LDDE). 
In Sect. \ref{sec:zshell_llclass}, we take a closer look at detailed behavior of the XLF in 
redshift divided shells and the number density/emissivity evolution in divided in luminosity classes. We also 
present binned XLFs. The results are discussed in Sect. \ref{sec:disc} and a concluding summary is given 
in Sect. \ref{sec:conc}. 

Throughout this work we use a Hubble constant $H_0=70 h_{70}$ ${\rm km\,s^{-1}\,Mpc^{-1}}$ $\Omega_{\rm m}=0.3$, and 
$\Omega_\Lambda=0.7$. The $h_{70}$ dependence is explicitly stated. If units are omitted, X-ray luminosities 
($L$ with any sub/super scripts) are measured in units of $h_{70}^{-2}{\rm erg\,s^{-1}}$, and column densities 
$N_{\rm H}$ in ${\rm cm^{-2}}$. The symbol $\log$ signifies a base-10 logarithm and $\ln$ a natural logarithm. 


\section{The COSMOS Sample}
\label{sec:cosmos}
\subsection{The {\sl XMM}-COSMOS Sample}

 As a part of the COSMOS collaboration, the {\sl XMM}-COSMOS program \citep[PI=G. Hasinger][]{hasinger07} 
included observations the entire 2.14 deg$^2$ COSMOS field with the EPIC camera on board the {\sl XMM-Newton}    
observatory over three observing cycles, AO-3, 4 and 6.  X-ray point source catalogs have been produced
using a sophisticated source detection and characterization procedure \citep{capelluti07,capelluti09}. The procedure
was developed as a part of the XMM-SAS package\footnote{\url{http://xmm.esac.esa.int/sas/}}. In the combined dataset, which 
is composed of 55 XMM-Newton observations with a total exposure of $\sim$1.5 Ms, 1887 unique X-ray sources have been 
detected. Optical counterparts and their multiwavelength properties were published for the X-ray source detected in 
the first 12 XMM-Newton observations by \citet{brusa07} and all 53 successful observations made before 2007 by 
\citet{brusa10} respectively. 


 In this work, we use X-ray sources detected in the hard (EPIC energy channel between 2 and 8 keV) band
in the 53 XMM-COSMOS observation data with a maximum-likelihood of $ML>10$, which corresponds to the false detection 
probability of $e^{-10}\approx 4.5\times 10^{-5}$. \citet{brusa10} published a multiwavelength catalog for the 
53-field XMM-COSMOS sources with a primary 2-10 keV flux cut of $S_{\rm X}\geq 3\times 10^{-15}\,{\rm erg\,s^{-1}\,cm^{-2}}$,
which is our {\sl XMM}-COSMOS base catalog. After this flux cut, we are left with  923 hard-band selected {\sl XMM}-COSMOS sources. 
We use the EPIC 2--8 keV countrate to 2--10 keV flux energy conversion factor (hereafter referred to as ECF) 
obtained by assuming a power-law spectrum with photon index $\Gamma=1.7$ absorbed by a Galactic absorbing column 
density of $N_{\rm H}=2.5\times 10^{20}{\rm cm^{-2}}$ \citep{capelluti09}. 

\subsection{The {\sl Chandra}-COSMOS Sample}

 With its unprecedented spatial resolution, the {\sl Chandra} X-ray observatory (CXO) provides superb
point source sensitivity in deep surveys. The {\sl Chandra} COSMOS program \citep[C-COSMOS, PI=M.Elvis;][]{elvis09}
is a large Chandra program with a total exposure of $\sim 1.8$ Ms with 49 observations with the ACIS-I instrument.
The observed fields are arranged in 7$\times$7 overlapping tiles, covering a central 0.9 deg$^2$ of the
COSMOS field. Inclusion of the information obtained by the recent {\sl Chandra} Legacy COSMOS survey (2.5 Ms, PI=F. Civano)
will be a topic of a future work. 
 A total of 1781 unique X-ray sources have been detected with a likelihood threshold of
$ML>10.8$ in any of the full (0.5--7keV), soft (0.5--2 keV) or hard (2--7 keV) bands, out of which 1017 are 
hard band detected \citep{elvis09,puccetti09}. 
 
 In \citet{elvis09}, the 2--10 keV flux is given for an ECF calculated assuming a $\Gamma=1.4$ power-law. 
To treat the sample in the same way as XMM-COSMOS, we have converted the 2--10 keV flux using the 
ECF assuming $\Gamma=1.7$ for our further analysis.  Unlike the case of XMM-COSMOS, \citet{civano12} did 
not impose any further flux limit in their multiwavelength identification catalog. However, we have imposed a 2-10 keV
flux ($\Gamma=1.7$) limit of $S_{\rm X}>1.6\times 10^{-15}$ ${\rm erg\,s^{-1}cm^{-2}}$, below which the $\log N-\log S$ curve
spuriously rises due to the Eddington bias. This removes the 24 faintest hard-band detected sources.  
Note that the $\log N-\log S$ curve in Fig. 9 of \citet{elvis09} effectively plots down to this flux level.
This leaves 993 C-COSMOS hard-band sources.     

         
\subsection{The Combined {\sl XMM-Chandra} COSMOS Sample}
\subsubsection{Sensitivity map combination}

\begin{figure}[h]
\begin{center}
\resizebox{!}{!}{\includegraphics[width=0.9\hsize]{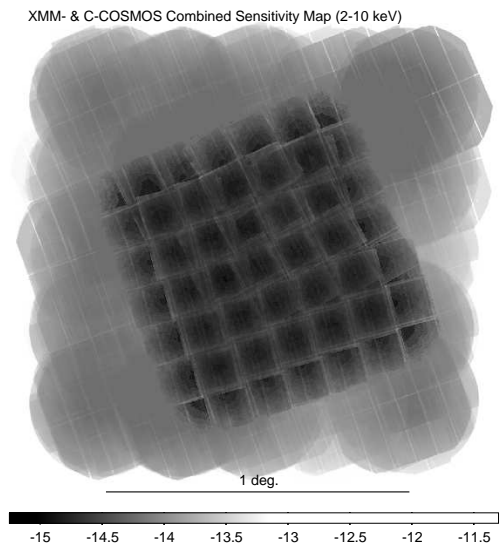}}
\end{center}
 \caption{The XMM-COSMOS/C-COSMOS combined sensitivity map. The scale indicated by a solid line 
   corresponds to 1 degree on the sky. The numbers at the tick marks 
   in the gray scale bar at the bottom of the figure indicate the base-10 logarithm of 
   the limiting 2-10 keV flux in units of ${\rm erg\,s^{-1}\,cm^{-2}}$ assuming a 
   $\Gamma=1.7$ power-law.
 } 

\label{fig:cosmossens}
\end{figure}

 In order to take advantage of both the XMM and C-COSMOS surveys, we have made a combined 
sample, which we call the XC-COSMOS sample, by selecting X-ray sources from the more sensitive 
survey out of the two at any given source position.  For this purpose, we have compared the hard X-ray sensitivity map 
for the 53-field XMM-COSMOS survey \citep{capelluti09} with that for C-COSMOS \citep{puccetti09}. The XMM-COSMOS 
sensitivity map has been truncated at the lowest limiting flux of the optical catalog described above.  
In order to compare both maps on the same grounds, we scaled the C-COSMOS sensitivity map 
using the ECF for the photon index of $\Gamma=1.7$. We further smoothed the C-COSMOS sensitivity map
with a Gaussian filter with $\sigma=16\arcsec$ and combined with the XMM-COSMOS
sensitivity map by choosing the smaller limiting flux at each point. We have also
generated a mask file that indicates whether the XMM-COSMOS or C-COSMOS source should 
used at each position. In effect, the C-COSMOS sources should be used in most of the region 
covered by C-COSMOS, except for small areas near corners and edges. The combined sensitivity map is shown in
Fig. \ref{fig:cosmossens}. In the combined sample, there are 1400 extragalactic point sources (mostly AGNs), 
18 Stars, and 29 unidentified sources. Overall the fraction of sources identified with AGNs/galaxies with spectroscopic 
or photometric redshifts and stars is 98\%. Note that the XC-COSMOS AGNs comprises
as many as $\sim 43\%$ of all AGNs in our global AGN sample defined below.   

 One of the greatest advantage of the COSMOS survey is the availability of multi-epoch, homogenized, 
deep 31 band photometry, covering from the UV to the MIR wavelengths, including 12 intermediate bands from 
Subaru \citep{taniguchi07}. These data, along with the correction for variability, and the use of hybrid AGN templates, 
have allowed the COSMOS survey to reach a mean photometric redshift accuracy 
of $\sigma_{\Delta z/(1+z_{\rm spec})}\approx 0.015$ with a low fraction of outliers ($\sim$5\%) for 
both the {\sl XMM} and {\sl Chandra} COSMOS samples \citep{salvato09,salvato11}.
  In this work, the newest photometric redshifts by \citet{salvato11} have been used when there 
is no available spectroscopic redshift. The associated probability density distributions of photometric redshifts 
are also used. In addition to the spectroscopic redshifts included in \citet{brusa10}, newly available proprietary spectroscopic
redshifts from spectroscopic survey programs within the COSMOS consortium have been included.
 
\begin{figure}[h]
\begin{center}
 \resizebox{!}{!}{\includegraphics[width=0.9\hsize]{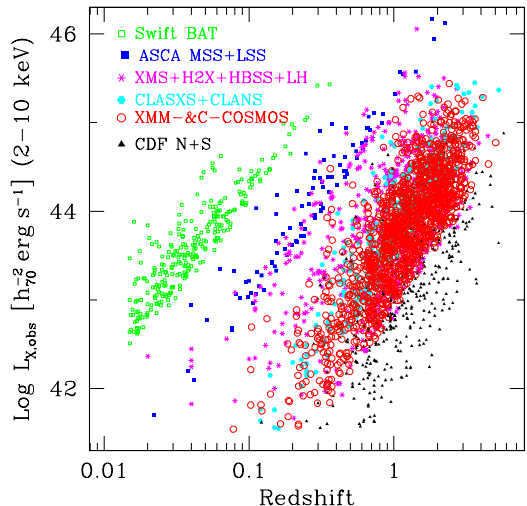}}
\end{center}
 \caption{The sample used in our analysis in the redshift--Observed X-ray luminosity space. 
 The surveys (or a group of similar surveys) are shown in different colors/symbols as labeled.
 (Color version of this figure is available in electronic version only.)}
 \label{fig:z_lgl}
\end{figure}

\section{The Global AGN Sample}
\label{sec:global_sample}

\subsection{The {\sl Swift} BAT Survey Sample}

 Based on the {\em Swift} BAT slew scan data accumulated over 60 months, \citet{ajello12} produced a catalog of AGNs
detected in the 15--55 keV band, providing the current most complete census of nearby bright AGNs over the entire sky.  
These include unabsorbed and absorbed AGNs up to moderate CTK AGNs. 
For CTN AGNs, 
photoelectric absorption affects very little to the 15--55 keV flux and thus there is no need for complicated absorption 
corrections. However, since we are evaluating intrinsic XLF at 2--10 keV, we need spectral template for unabsorbed AGNs to 
convert from 15--55 keV to 2--10 keV. In our simplest treatment, we assume a power-law spectrum with a photon index of 
$\Gamma=1.7$, which is approximately (within $\sim 4\%$) equal to the corresponding ratio for the unabsorbed AGN template 
spectrum (see Sect. \ref{sec:xlf_abs}). Under this assumption, we can convert the 15--55 keV flux to 2--10 keV flux by 
multiplying by a factor of 0.705. In the limiting flux-survey area curve from Fig. 1 of \citet{ajello12}, the flux 
is converted to the 2--10 keV value in this way. In our full treatment in Sect. \ref{sec:xlf_abs}, we use 
a luminosity-dependent effective photon index for the conversion.
        
\subsection{{\sl ASCA} LSS/MSS}

 The Advanced Satellite for Cosmology and Astrophysics ({\sl ASCA}) was the first X-ray observatory
with focusing optics capable of making spectrocopic imaging in the 2-10 keV band.
The combination of {\sl ASCA} Large Sky Survey (ALSS) and the {\sl ASCA} Medium Sensitivity survey (AMSS)
fills the gap between all-sky surveys and deeper surveys with {\sl XMM-Newton} and {\sl Chandra}, 
occupying a unique regime in the $z-L_{\rm X}$ space at $E>2$ keV.

The ALSS covers a continuous area of 5.5 deg$^2$ near the North Galactic Pole with a flux limit of 
$S_{\rm X}\sim 1 \times 10^{-13}{\rm [erg\,s^{-1}\,cm^{-2}]}$ (2 -- 10 keV) \citep{ueda98,ueda99}. 
Thirty AGNs detected with the SIS instrument are completely identified by \citep{akiyama00}.
The AMSS is based on a serendipitous X-ray survey with the GIS instrument\citep{ueda01,ueda05} and 
has a survey area of $\sim$ 85 deg$^2$. The flux limit ranges from 
$5\times 10^{-12}$ to $3\times 10^{-13}$ ${\rm erg\,s^{-1}cm^{-2}}$. 
 The identification catalogs in the northern part of the AMSS \citep[][AMSSn;]{akiyama03} and 
the southern part (AMSSs; Akiyama et al. in prep.) contain 74 and 
20 spectroscopically identified non-blazer AGNs respectively.  Three X-ray sources are left unidentified and the identification
completeness is 97\% for the combined ALSS and AMSS sample.

 The 2-10 keV fluxes of these sources have been obtained from the {\sl ASCA} GIS count rate assuming a 
$\Gamma=1.7$ power-law using the GIS response matrix.  

\subsection{The {\sl XMM-Newton} Hard Bright Serendipitous Sample (HBSS)}

The Hard Bright Serendipitous Sample \citep[HBSS][]{dellaceca04} is a subsample detected in the 4.5 –- 7.5 keV 
band from the larger survey of the {\sl XMM-Newton} Bright Survey (XBS). \citet{dellaceca08} defined a complete
flux-limited sample of 67 sources with MOS2 count rates larger than 0.002 $[{\rm cts\,s^{-−1}]}$ 
(4.5 -- 7.5 keV) over 25 deg$^2$ of the sky. The MOS2 count rates in the HBSS were converted to observed 
fluxes in the 2-10 keV band, assuming a photon power-law index of 1.7 (using an ECF of 
$1.044 \times 10^{-11}~{\rm erg~cm}^{-2}~{\rm cts}^{-1}$. The corresponding flux limit is 
$2.1 \times 10^{-13}~{\rm erg~s^{-1}cm^{-2}}$. These sources are completely identified spectroscopically 
except for two, making the identification completeness of $\sim 97$\%. The 65 identified sources 
consist of 62 extragalactic point sources (mostly AGNs), 2 stars and 1 cluster of galaxies.

\subsection{The {\sl XMM-Newton} Medium-sensitivity Survey (XMS)}

The {\sl XMM-Newton} Medium-sensitivity Survey (XMS;\citealp{barcons07}) is a serendipitous X-ray survey and optical identification 
program of sources with intermediate X-ray fluxes discovered in 25 XMM-Newton high Galactic latitude fields 
covering a sky area of $\sim $3 deg$^2$. For this analysis the XMS-H sample, selected in the 2-10 keV band, is used.
\citet{barcons07} assumed a spectral index of $\Gamma=1.7$ for the count-rate to flux conversion for the 
hard band, and therefore we use the 2-10 keV flux as shown in their dataset.  In our analysis, we use the 
the sub-sample of XMS-H defined by \citet{hasinger08}. 
  
 The original \citet{barcons07} sample contains 159 sources with 2-10 keV fluxes $>3.3 \times 10^{-14} {\rm [erg\,cm^{-2}\,s^{-1}]}$ 
with a spectroscopic identification fraction of 83\% (27 unidentified sources). However, the actual spectroscopic 
completeness limit varies from field to field. Therefore \citet{hasinger08} selected a subsample that is comprised of 
further flux cuts on a field-by-field basis. This way a cleaner XMS sample could be defined, including 128 extragalactic point X-ray sources
and almost half the number of unidentified sources, i.e. achieving an identification fraction of 91\%. 
The survey solid angle has been corrected accordingly. See Sect. 2.4 of \citet{hasinger08} for further details.

\subsection{Hellas2XMM}

 HELLAS2XMM \citep{baldi02} is a serendipitous survey based on suitable XMM-Newton pointings (complementary to the XMS).
The assumed spectral index for the ECF for the 2-10 keV flux in \citet{baldi02} was $\Gamma=1.7$, and therefore we use their flux as provided.

 As in the case of the XMS, we use the sub-sample compiled by \citet{hasinger08}.  
\citet{fiore03} presented optical identifications and spectroscopic redshifts for 122 sources selected in the 
2-10 keV band in five XMM-Newton fields, covering a survey area of 0.9 deg$^2$. Later 
\citet{maiolino06} presented additional redshifts from VLT NIR spectroscopy of optically extremely faint objects for two 
additional sources. Both \citet{fiore03} and later \citet{mignoli04} tried to estimate redshifts for the remaining, 
optically faint, unidentified sources. \citet{cocchia07} published photometry and spectroscopic redshifts in five 
additional HELLAS2XMM fields providing 59 new redshift identifications for the sample of 110 new sources
with an additional solid angle of 0.48${\rm deg^2}$. As explained in \citet{hasinger08}, further flux cuts 
were made on the field-by-filed basis. In addition, different flux cuts were applied to inner and outer regions 
of each field. The redshift completeness in this subsample of HELLAS2XMM is thus 84\%. In this paper,
we further raise the flux limit from 1.0$\times 10^{-14}$ \citep[adopted by][]{hasinger08} 
to 1.63$\times 10^{-14}{\rm erg\,s^{-1}cm^{-2}}$ to make the identification 90\% complete, leaving 115 AGNs.  

\subsection{Lockman Hole ({\sl XMM})}

Because of the extremely low Galactic column density ($N_{\rm H}=5.7\times 10^{19}{\rm cm^{-2}}$), Lockman 
hole has been selected as a target of X-ray deep survey observations, including a {\sl XMM-Newton} survey 
\citep{hasinger01,brunner08}.
The base X-ray source catalog for this work is from \citet{brunner08} from 18 individual pointings, twelve of
which are centered near the same pointing direction and the remaining pointings are spread out over about 
$\sim 30\arcmin$ in right ascension. \citet{brunner08} concentrated on the 
deep central circular area with a radius of 15$\arcmin$ ($= 0.20 {\rm deg^{2}}$). Their catalog contains
409 X-ray sources, out of which 266 are detected in the hard (2-4.5 keV) band at the maximum-likelihood
(ML) value of 6.0 or above. Based on simulations, they showed that the sources above this ML threshold,
in combination with the flux-sensitivity curve in their Fig. 5, reproduces the $\log N-\log S$ accurately.   

 Their listed 2-10 flux is converted from the EPIC 2-4.5 keV count rate assuming a $\Gamma=2.0$
power-law spectrum. For our purposes, we have converted their $\Gamma=2.0$ based 2-10 keV flux to the $\Gamma=1.7$ based 
flux using a combined response matrix of the {\sl XMM-Newton} EPIC PN and two MOS detectors. This
corresponds to a multiplication of the $\Gamma=2.0$ based flux by 1.16.  The limiting-flux -- area curve in 
Fig. 5 of \citet{brunner08} has also been converted to the $\Gamma=1.7$-based flux. 

 Of the 266 hard X-ray selected sources, 60\% have published or unpublished spectroscopic redshifts (including  
5 stars). Including photometric redshifts by \citet{fotopoulou12}, 98\% are identified with either spectroscopic or 
photometric redshifts. Only 4 are left unidentified. This results in 257 hard-band selected extragalactic X-ray point sources. 
Probability density distributions of the photometric redshifts are also available, which we use in Sect. \ref{sec:pzpdf}.  
    
\subsection{CLASXS and CLANS}

 The catalogs from two  intermediate-depth {\sl Chandra} surveys near the Lockman Hole, the {\sl Chandra} Large Area Synoptic 
X-ray Survey (CLASXS; \citealp{yang04}) and the {\sl Chandra} Lockman Area North Survey (CLANS;\citealp{trouille08}) are also included 
in our analysis. The X-ray source identifications, and limiting-flux -- survey area curves of both surveys (as well as 
those of the CDF--N; see below) are conveniently included in \citet{trouille08} as a part of their
``Opt-X'' project. Additional spectroscopic redshifts of CLANS sources can also be found in \citet{trouille09}. Their tables 
contain 2-8 keV fluxes based on the 2-8 Chandra ACIS-I count rates with the ECFs calculated with 
variable $\Gamma$ and ACIS-I response matrices based on old calibrations. Therefore we have converted from their 2-8 keV 
count rate to 2-10 keV flux assuming $\Gamma=1.7$ and with ancillary response files (ARFs) we have re-created for the ACIS-I aim point
using the Chandra CALDB 4.5.7  (the latest calibration as of our analysis).  For each of the CLASXS and CLANS surveys, 
the applied response is the exposure weighted average of the responses at the aim points of the observations belonging
to the survey. The flux in their limiting flux--survey area curves \citep[Fig. 5 in ][]{trouille08} have been 
adjusted accordingly. The curves for the probability of detection of 30\% are used, because they reproduce 
the $\log N - \log S$ relation in the simulations by \citet{yang04}.  Because their curves only apply to the sources with 
offaxis angles of $<8\arcmin$, we use sources that meet this criterion. Also, we only include X-ray sources that are 
detected at a signal-to-noise ratio of $>3$ in the hard band in order to be compatible with their flux--survey area curve. 
Moreover, in order to be compatible with there flux-limit -- area curve, they also imposed offaxis-dependent flux limits to exclude
some sources for their $\log N - \log S$ calculations. These imposed limits cannot be reproduced with the published information. 
However, this effect does not significantly affect our analysis as described below.

 In order to minimize the effects of the ``completeness correction'' (Sect. \ref{sec:completeness}), we impose at least 90\% identification 
completeness for each survey.  In order to achieve this, we have
excluded the sources with 2-10 keV flux ($\Gamma=1.7$ based) of $S_{\rm X}<6.5 \times 10^{-15}$ and  
$S<6.5 \times 10^{-15}$ ${\rm erg\,s^{-1}cm^{-2}}$, leaving 250 and 119 AGNs  for CLASXS and CLANS respectively.  
At these flux limits, the survey area is already $\sim 95\%$ of the total geometric area with offaxis angle $<8\arcmin$. 
Thus the effects of offaxis-dependent flux limits are minimal. 

\subsection{{\em Chandra} Deep Field North (CDF--N)}

 Our base X-ray source catalog for the 2Ms CDF--N is from  \citet{alexander03}. As in the cases
of CLASXS and CLANS, we have used the hard-band detected sources that have signal-to-noise ratio of 3 or better
to match with the flux-limit survey area curve shown in their Fig. 19. In order to convert their 2-8 keV count rate 
to 2-10 keV flux, we have re-generated the ancillary response file for an exposure-weighted 
average of all the CDF-N Chandra observations used by \citet{alexander03}, assuming a $\Gamma=1.7$ 
power-law and using CALDB 4.5.7. Since their flux-limit survey area curve is based on  $\Gamma=1.4$ and old calibration, we
have adjusted their curve for our flux definition. 

 We use the optical identifications listed in \citet{trouille08}, which include redshifts from the
literature as well as from their own program. We have further imposed a lower
flux cut of $S_{\rm X}>8.9\times10^{-16}$ ${\rm erg\,s^{-1}\,cm^{-2}}$ to ensure an identification 
completeness of at least 90\%, leaving 182 extragalactic X-ray point sources.

\subsection{{\em Chandra} Deep Field South (CDF--S)}\label{sec:cdfs}

 An extensive X-ray source and identification catalog of the 4 Ms Chandra Deep Field South was published by 
\citet{xue11}, which also includes a list of previously published spectroscopic and photometric redshifts. 
Subsequently \citet{lehmer12} generated limiting flux -- solid angle curves and the $\log N-\log S$ relation.
This work took into account flux probability distributions and recovery functions for those sources  
with a threshold probability of false detection of $P_{\rm thresh}=0.004$, as adopted by \citet{xue11}.

 We have chosen hard X-ray sources that have been detected in the 2-8 keV band from their catalog. 
Their count-rate to flux conversion is based on the effective photon index $\Gamma$ from 
the ratio of the hard and soft band count rates, and therefore the ECF is variable.
For each source, we have converted their 2-8 keV flux (based on their estimated $\Gamma$) to 2-10 keV flux
based on $\Gamma=1.7$ using the aim point response matrix of one of the CDF--S observations (OBSID=12049).
As shown below, the value of the count-rate to flux conversion factor is subject to the variation
of instrumental response over time as well as calibration updates, but the relative change of 
conversion factors among different spectra is insensitive to ACIS-I response matrix used. 
The  2-10 keV flux (after the conversion) of the faintest object amongst the hard-band detected sources 
was 8.3$\times 10^{-17} {\rm erg\,s^{-1}cm^{-2}}$. 
     
 We have used the 2-8 keV band limiting flux -- solid angle curve in Fig. 2 of \citet{lehmer12}. Their
curves are drawn for $\Gamma=1.0$ and we have converted their relation to $\Gamma=1.7$ using the 
method described above. 

 We have further included new spectroscopic redshifts that are not included in the \citet{xue11} catalog.
In the catalog of \citet{xue11}, there are two objects with photometric redshifts above $z_{\rm ph}>7$.
The second solution of these objects by \citet{luo10} are both at $z_{\rm ph}\approx 3$. However,
a new spectrocopic redshift obtained for one of them (XID=28) is $z_{\rm sp}=0.69$. Thus it is possible
that the other also suffer a similar catastrophic photometric redshift error also. Thus we have 
excluded the remaining $z_{\rm ph}>7$ object from our sample. The overall identification completeness
is 97\% (including photometric redshifts) and the spectrocopy completeness is 71\%.

\subsection{Removing Compton-thick AGNs}

{ The 2-10 keV selected samples, especially at low redshifts, are highly selected against CTK AGNs. Even 
the {\it Swift} BAT sample selected at 15-55 keV suffers from suppression of the CTK population, because of  the Compton 
scattering itself which results in subsequent lowering of the photon energy as well as longer paths in absorbing medium. 
However, there always is some  spill over of the CTK population to the sample defined here. There are two approaches in treating the situation.  
One is to include the CTK population into the model and make model fits to the sample that may include
CTK AGNs. In other words, the amount of CTK AGN spill-over is estimated by the model.  
The model of U14 introduced the parameter $f_{\rm CTK}$, which is the ratio of the number density of the CTK AGNs 
($24\leq \log N_{\rm H}<26$) to that of absorbed CTN AGNs ($22<\log N_{\rm H}<24$), and they then assumed the same $f_{\rm CTK}$ at
all luminosities and redshifts. They showed that this is at least consistent with various observational 
estimates of the CTK AGN population in the literature. However, U14 also estimated that the observational constraint on this 
parameter was $0.5<f_{\rm CTK}<1.6$, carrying a factor of three uncertainty. This is consistent with recent results by 
\citet{buchner15}. An alternative approach is to remove known 
CTK AGNs from the sample on a best effort basis and construct an XLF model that represents only the CTN AGNs. 
We take the latter approach.} 

{ There are a number of studies aimed at  identifying CTK AGNs among the AGN samples used in this paper. 
 Because the {\sl Swift} BAT sample is selected in the 15-55 keV band, it has some sensitivity to  CTK AGNs with 
$\log N_{\rm H}\la 25$. \citet{ajello12} list 18 known CTK AGNs from their preceding work of 
\citet{burlon11} and other literature. These have been excluded from our analysis. \citet{ajello12} estimated that 
a few {\em unknown} CTK AGNs might still be present in this sample, the effect of which on our analysis is negligible. 
In their {\sl XMM-Newton} Lockman hole sample, 11 AGNs meet the X-ray color-color criteria of CTK AGNs defined by \citet{brunner08}. 
None of them is in our hard-band selected sample above our luminosity cut. \citet{lanzuisi15} selected 10 CTK AGNs based 
on spectral fits for the XMM-COSMOS sources that have 30 counts or more in the full 0.3-10 keV band. None of them are in the {\sl XMM} 
part of the  XC-COSMOS AGN sample. Note 
that all but one XMM-Newton AGNs in the XMM part of the XC-COSMOS sample have 0.5-10 keV counts greater than 30, thus, 
\citet{lanzuisi15} would have identified almost all CTK AGNs in it if any were present. One AGN in our CLANS sample is 
among the five X-ray selected CTK AGNs by \citet{polletta07} and this has been excluded for our analysis. Nine of   
the CDF-N AGNs are among the CTK AGN candidates derived from the X-ray spectral analysis by \citet{georgan09}. These are also
excluded from our analysis. \citet{georgan09} performed spectral analysis of the CDF-N AGNs with a 2-10 keV flux of $>1\times 10^{-15}$
$[{\rm erg\,s^{-1}\,cm^{-2}}]$, which is approximately the same as our flux cut to the CDF-N sources to meet our 
completeness criteria. Recently \citet{brightman14} published a table of extensive X-ray spectral analysis results 
of the X-ray sources in  the CDF-S, AEGIS-XD and C-COSMOS surveys with an intention to identify CTK AGNs.  Because of the low background level of {\sl Chandra}
surveys, they attempted spectral analysis to the sources down to 10 {\sl Chandra} ACIS counts. Following the criteria 
by \citet{brightman14}, we consider as probable CTK AGNs those that satisfy the both of the conditions; 
the best-fit value of $\log N_{\rm H}$ is greater than $>24$ and the lower bound of its 90\% confidence error range 
is $>23.5$.  We exclude these from from our analysis also. The number of excluded sources by the CTK AGN criteria are 24 
for the C-COSMOS and 23 for the CDF-S samples respectively. The full band counts of almost all faintest sources above our 
flux cuts (see above) well exceed 10 counts for both C-COSMOS and CDF-S samples. Thus CTK AGNs that may be still remain in 
our sample can be neglected. The CTK AGNs are still a minor population in our sample, even at the faint end.  The CTK AGNs identified above 
comprise  $19\pm 6\%$, $9\pm 3\%$, $3.6\pm 0.5\%$, and $1.9\pm 0.3\%$ for the $\log S_{\rm x}$ ranges of 
$(-16.1,-15.5)$, $(-15.5, -15.0)$,  $(-15.0,-14.5)$, and $(-14.5,-14.0)$ respectively, where the value of -16.1 corresponds to
the faintest source in our sample. These fractions are consistent with the model prediction by U14 for $f_{\rm CTK}=1$ 
(see Fig. 17 of U14).}

\subsection{Sample Summary}
\label{sec:global_samp}

 In each sample except for {\sl Swift} BAT, we have converted the 2--10 keV fluxes of all the sources using the energy 
conversion factor (ECF) assuming a power-law spectrum with a photon index of $\Gamma=1.7$. 
To represent fluxes calculated in this way, we use the symbol $S_{\rm X,obs}$, or $S_{\rm X,obs}^{\Gamma=1.7}$ 
when we want to emphasize that we use the ECF based on $\Gamma=1.7$. We also define the {\em observed luminosity} 
\begin{equation}
L_{\rm X,obs}\equiv 4\pi d_{\rm L}(z)^2 S_{\rm X,obs},
\label{eq:lobs}
\end{equation}  
where $d_{\rm L}(z)$ is the luminosity distance at the redshift $z$. Implementation of the K-corrections and absorption 
corrections using realistic AGN spectra is discussed in Sect. \ref{sec:xlf_abs}. In our further analysis, we are interested 
in X-ray sources with intrinsic luminosities of $\log L_{\rm X}> 42.0$ and we use, in effect, X-ray point sources with 
$\log L_{\rm X,obs}\ga 41.5$. Thus our analysis is strongly de-selecting non-AGN X-ray sources, the number density of which 
is lower than AGNs by at least an order of magnitude at $\log L_{\rm X}\ga 41.5$ \citep{persic04,ranalli05,ptak07}. Our 
sample is plotted in the $z-\log L_{\rm X,obs}$ space in Fig. \ref{fig:z_lgl}.  

\begin{deluxetable}{lccccc}
\tabletypesize{\footnotesize}
\tablecaption{The X--ray samples\label{tab:samp}}
\tablewidth{0pt}
\tablehead{
 \colhead{Survey} & \colhead{Max. Solid Angle} &  \colhead{$\log S_{\rm X,obs}^{\rm lim}$\tablenotemark{a}} & 
 \colhead{$N_{\rm sz}$\tablenotemark{b}}  & \colhead{$N_{\rm pz}$\tablenotemark{c}} & \colhead{Completeness}\tablenotemark{d} \\
 \colhead{} & \colhead{[deg$^2$]}  & \colhead{[cgs]} & \colhead{} & \colhead{} & \colhead{\%}  
}
\startdata
{\sl Swift} BAT    & $3.9\times 10^4$  &  -11.4  & 274    & 0 & 98   \\
{\sl ASCA} LSS+MSS &      $91.$        &  -12.5  &  94    & 0 & 97  \\
 HBSS              &      $25.$        &  -12.6  &  62    & 0 & 97  \\
 Hellas2XMM        &      $3.8$        &  -13.5  & 115   & 0 &  90  \\
 XMS                  &     $1.38$        &  -13.7  & 128 & 0 &  91  \\
 CLANS                &     $0.49$        &  -14.5  & 183 & 63  & 90 \\
 CLASXS               &     $0.32$        &  -14.4  &  96 & 23  & 90 \\
 XC-COSMOS            &     $2.17$        &  -14.6  & 940 & 424 & 98 \\
 LH-XMM               &     $0.20$        &  -14.7  & 155 & 102 & 98 \\ 
 CDF-N                &    $0.125$        &  -15.5  & 118 & 54  &  90 \\
 CDF-S                &    $0.128$        &  -15.8  & 254 & 85  &  97 \\
\enddata
\tablenotetext{a}{The limiting flux corresponding to 20\% of the maximum solid angle of each survey. 
  The fluxes are in the 2-10 keV band.}
\tablenotetext{b}{The number of extragalactic X-ray point sources with spectroscopic redshifts.}
\tablenotetext{c}{The number of extragalactic X-ray point sources with photometric redshifts, but with no spectroscopic redshifts.}
\tablenotetext{d}{Identification completeness.}
\end{deluxetable}


{ The properties of each sample used in our analysis is summarized in Table \ref{tab:samp}, where the name  
of the survey, the survey solid angle, the limiting flux corresponding to 20\% of the maximum solid angle, the number
of extragalactic point sources with spectroscopic redshift, those with photometric redshift, and identification completeness
are listed for each survey. The numbers in this table do not include CTK AGNs that have been discussed in the previous section nor 
those below our luminosity cut.} 

\subsection{Incompleteness Correction and Number counts}
\label{sec:completeness}

\begin{figure*}
   \begin{center}
   \includegraphics[width=0.48\hsize]{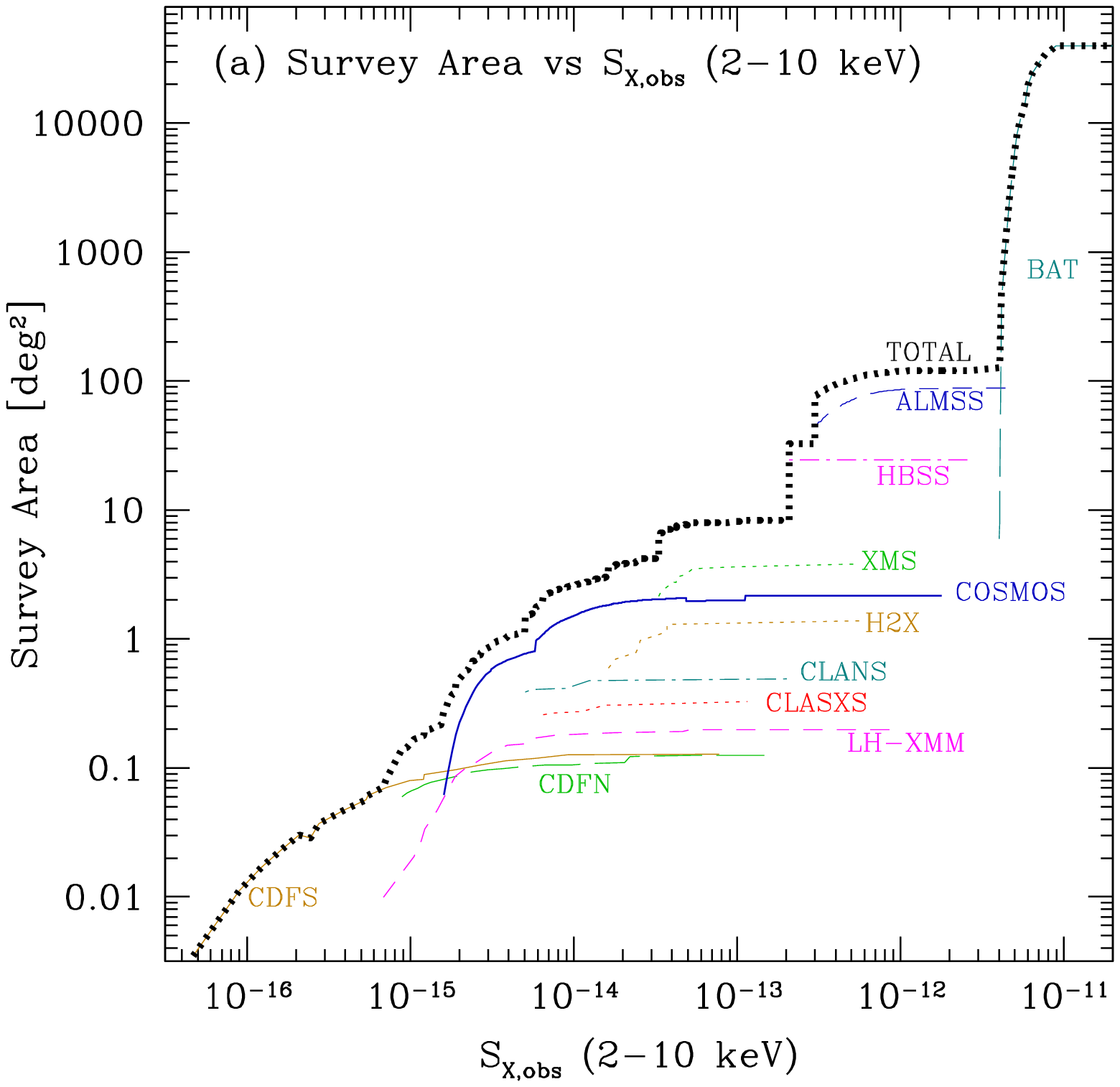}
   \includegraphics[width=0.48\hsize]{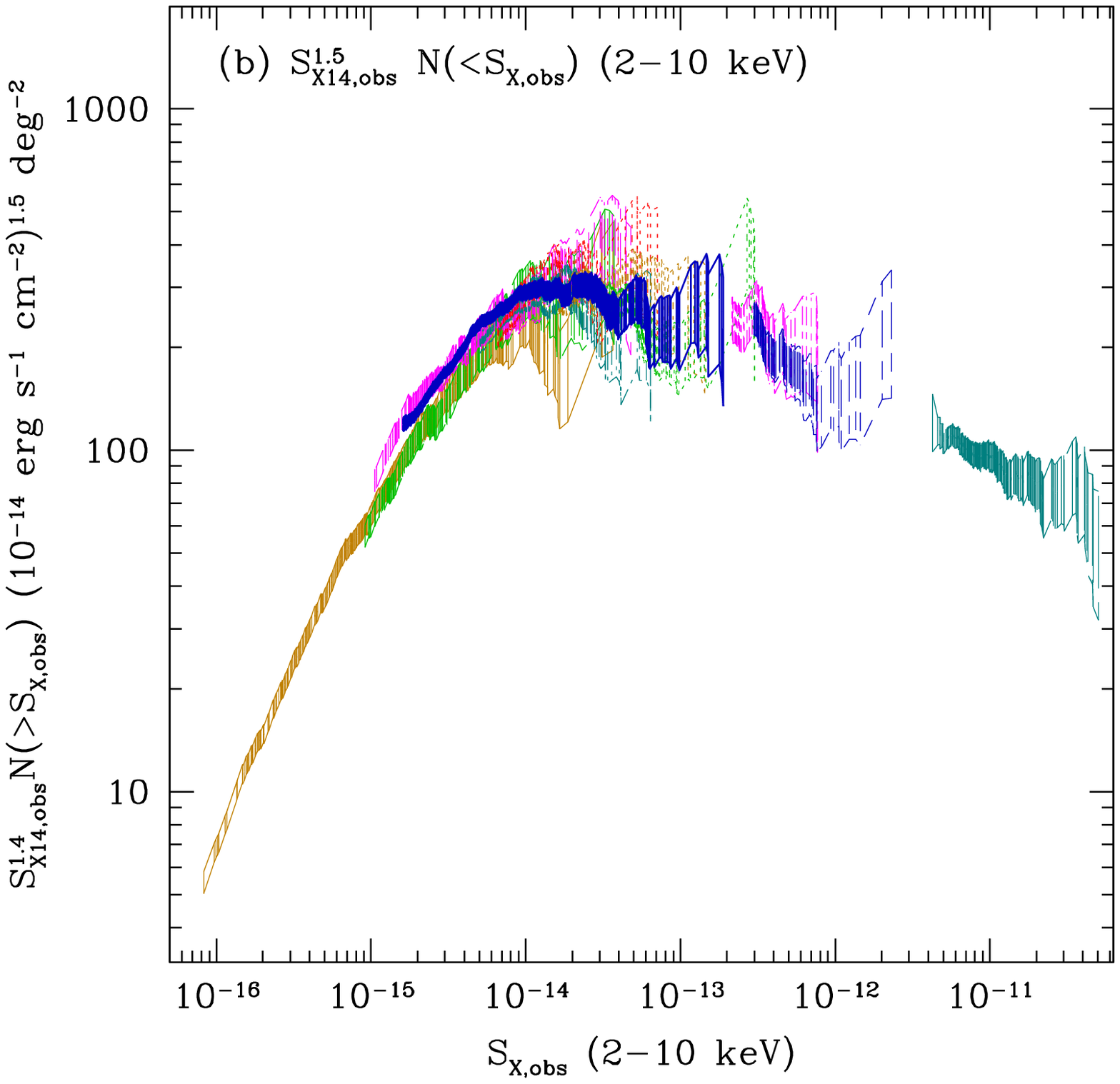}
   \end{center}
   \caption{
     (a) The completeness-corrected survey area curve as a function of 
     the 2-10 keV flux limit. A photon index of $\Gamma=1.7$ is assumed for all
     count-rate to flux conversions. The total area as well as those for individual 
     surveys are shown as labeled. 
     (b) The ``tilted'' cumulative number counts $S_{X14,obs}^{1.5}N(>S_{\rm X,obs})$ for individual samples with 1$\sigma$ errors.  
     The multiplication factor $S_{\rm X14,obs}^{1.5}$ is applied to the cumulative number to make the Euclidean slope flat.  The meaning of line 
     colors and styles corresponds to those in the the flux limit -- survey area curves in panel (a). For visibility,
     the data points corresponding to the 4 brightest objects in each sample are not displayed. (Color versions of these figures
     are available in the electronic version.)
   }
   \label{fig:area_ns} 
\end{figure*}

 While the identification completeness of our selected samples are at least 90\%, we further account for unidentified 
sources as follows. For samples with five or less unidentified sources, we have multiplied the survey area in the flux-limit --- 
survey area curve by the identification completeness of the sample over the whole flux range, assuming that the redshift (or 
the source type) probability distribution of an unidentified source is the same as the redshift/source type distribution of 
the identified sources in the same sample. 

 If there are more than five unidentified sources, we have made a similar adjustment of
the survey area in flux blocks. We divide the X-ray sources into flux blocks from
the brightest to faintest, in such a way that a new block starts after the
fifth unidentified source.  The last (faintest) block may contain one to five unidentified 
sources. For each block, we have adjusted the survey area by multiplying by the completeness
level of the block. The assumption is that the redshift/source type probability distribution 
is the same as that of the identified source at similar X-ray fluxes.   

 The resulting incompleteness-corrected limiting-flux curve of each survey as well as that for the total
of all surveys are plotted in Fig. \ref{fig:area_ns}(a). In order to verify the integrity and consistency 
of our samples and the level of systematic errors and/or the effects of cosmic variance, 
we plot the ``tilted'' cumulative number counts $S_{\rm X14,obs}^{1.5}\,N(>S_{\rm X})$ curves in Fig. \ref{fig:area_ns}(b),
where $S_{\rm X14,obs}$ is the 2-10 keV observed flux (see above) measured in units of $10^{-14}{\rm erg\,s^{-1}cm^{-2}}$, 
for each of our identified extragalactic X-ray point source sample.  The cumulative count is
calculated by:
\begin{equation}
N(>S_{\rm X,obs}) = \sum_{S_{{\rm X,obs},i}>S_{\rm X}}1/\Omega_{\rm c}(S_{{\rm X,obs},i}),
\end{equation}
where $\Omega_{\rm c}(S_{{\rm X,obs},i})$ is the corrected survey area at the limiting flux 
of the $i$-th source $S_{{\rm X,obs},i}$. One sigma errors, calculated by
\begin{equation}
\sigma\left[N(>S_{\rm X,obs})\right]^2 = \sum_{S_{{\rm X,obs},i}>S_{\rm X,obs}}\left[1/\Omega_{\rm c}(S_{{\rm X,obs},i})\right]^2,
\end{equation}
are shown. For visibility, we do not plot data points corresponding to the four brightest object of the survey.
 
\section{The XLF calculations}
\label{sec:xlf_calc}

\subsection{Parametric Modeling with maximum likelihood fitting}
\label{sec:ml12}

We follow the same procedure as our previous work \citep{miyaji00a,miyaji01,ueda03,ueda14} and determine the 
best-fit parameters of parameterized model of the XLF. 

 As our maximum-likelihood estimator, we use either one of the following two forms:  
\begin{eqnarray}
&\mathcal{L}=-2\;\sum_i \ln \left[N(\log\;L_{{\rm X}i},z_i)\right]+\nonumber \\
& \hspace{2em} 2\int\int N(\log\;L_{{\rm X}},z){\rm d}\log\;L_{\rm X}\;{\rm d}z\label{eq:ml1}\\
&\mathcal{L}=-2\sum_i\ln\left[\frac{N(\log\;L_{{\rm X}i},z_i)}
 	{\int \int N(\log\;L_{{\rm X}},z){\rm d}\log\;L_{\rm X}\,{\rm d}z}\right],\label{eq:ml2}
\end{eqnarray} 
where $i$ goes through each AGN in the sample and $N(\log\;L_{{\rm X}},z)$ (N-function) is the expected number 
density of AGNs in the sample per logarithmic luminosity per redshift, calculated from a parameterized 
analytic model of the XLF:

\begin{eqnarray}
       \nonumber \\
N(\log\;L_{{\rm X}},z) = &
	\nonumber \\
\frac{{\rm d}\;\Phi}
	{{\rm d\; Log}\;L_{\rm X}}\;d_{\rm A}(z)^2\;(1+z)^2 & 
        \frac{{\rm d}\chi}{{\rm d}z}(z) \cdot A(S_{\rm X,obs}),
\label{eq:n_func}
\end{eqnarray} 
where $d_{\rm A}(z)$ is the angular diameter distance, $\chi$ is the radial 
comoving distance \citep[e.g., Sect 4.3 of][]{schneider06},
and $A(S_{\rm X,obs})$ is the survey area as a function of limiting observed 
X-ray flux $S_{\rm X,obs}$ as shown in Fig. \ref{fig:area_ns}(a). For a power-law spectrum with a 
photon index of $\Gamma$, 
\begin{equation}
S_{\rm X,obs}=L_{\rm X}/4\pi d_{\rm L}(z)^2(1+z)^{2-\Gamma},
\label{eq:sxobs}
\end{equation}
$d_{\rm L}(z)$ is the luminosity distance.

 The minimization of  $\mathcal{L}$ with respect to model parameters gives the best-fit model. With Eq. \ref{eq:ml2},
employed in our previous works, the parameter that represents the global normalization cannot be a fitting 
parameter, since within the estimator itself, the model number density is normalized. 
On the other hand, with Eq. \ref{eq:ml1}, employed by \citet[e.g.,][]{marshall83}, one can treat 
the normalization as a fitting parameter.      
 
 We use the MINUIT package \citep{minuit} distributed as a part of the CERN program library
for the minimization procedure and parameter error search. The MINUIT command ``MINOS'' gives errors taking parameter 
correlations into account. In cases of global expressions (Sect. \ref{sec:xlf_global}), ``MINOS'' fails to give errors 
for many parameters if we use Eq. \ref{eq:ml1}. In this case, we use Eq. \ref{eq:ml2} and the normalization and its errors 
are calculated independent of the fitting process as follows. The model normalization $A$ (as a generic symbol 
of various normalizations that appears in the following subsections) is determined, such that the total number 
of expected objects is equal to the number of AGNs in the sample ($N^{\rm obs}$). The estimated 1$\sigma$ confidence error 
for $A$ is taken to be $A/\sqrt{N^{\rm obs}}$ and does not include correlations of errors with other parameters. In our results, 
we indicate which method has been used for each fit.

\subsection{Photometric redshift full probability density distribution}
\label{sec:pzpdf}

 For the CLASXS, CLANS, COSMOS, LH, and {\sl Chandra} Deep fields North and South datasets, significant fractions of 
redshifts rely on photometric redshift determinations. Especially for fainter optical sources, which are systematically
selected against for spectroscopic measurements, the error in the primary peak and catastrophic failure rate 
increase rapidly. Thus we include the probability density distribution (PDF) in redshift space for photometric 
redshifts, whenever available.

 Let the PDF of the i-th object be $p_i(z)$ (normalized with $\int p(z)dz=1$). The X-ray luminosity of the $i-th$ object
$L_{\rm x,i}(z)$ is now a function of $z$. The PDFs can be incorporated into Eqs. \ref{eq:ml1} \& \ref{eq:ml2}  
by replacing $N(\log\;L_{{\rm x}i},z_i)$ by: 

\begin{equation}
 \int N(\log\;L_{{\rm x}i}(z),z))p_i(z)dz.
\end{equation}  

 We use the  PDFs derived by \citet{salvato11} and \citet{fotopoulou12} for the COSMOS and 
LH samples respectively. They give the PDFs in bins of $\Delta z=0.01$ in the $0<z<7$ range. To reduce the computational time and at the same
time not to sacrifice the accuracy, we rebinned the PDFs into $\Delta z/(1+z)\approx 0.03$ bins and then neglected the new bins 
that have a probability of less than 2\%.  For the CLASXS, CLANS, and CDF samples, no PDF information was available in the public 
domain at the time of our analysis. However, for the CDF-S, \citet{xue11} show the first and second peaks of the photometric redshift 
PDF derived by \citet{luo10} for some of their photometric redshifts. For those, we assumed that the PDF is the sum of two delta functions 
centered at these peaks, each of which has a weight of 0.5. The effects of considering the PDFs are discussed in Sect. \ref{sec:pdf_spec}.

\subsection{Using Realistic AGN Spectra}
\label{sec:xlf_abs}

 Our sample contains AGNs at various absorption levels. Thus we have considered realistic absorbed spectra
in our XLF calculations, following the approach by U14 \citep[see also][]{miyaji00b,ueda03,ebrero09}. In this work, 
we do not estimate the $N_{\rm H}$ on individual AGNs nor derive the $N_{\rm H}$ function from our own dataset. 
Instead, we include in our model fitting the latest luminosity and redshift dependent $N_{\rm H}$ function,
$f(L_{\rm X},z;\log N_{\rm H})$, from the literature and AGN template spectrum. \citet{ueda14}
derived the refined $N_{\rm H}$ function and its evolution, where the evolution of absorbed AGN fraction is based on  
that of \citet{hasinger08}. Writing down the full form of the $N_{\rm H}$ function is beyond the scope of this paper and the 
reader is directed to citet{ueda14}. The $N_{\rm H}$ function is normalized as 
$\int_{20}^{24}f(L_{\rm X},z;\log N_{\rm H})d\log N_{\rm H}=1$ 
(unabsorbed AGNs with $\log N_{\rm H}<20.0$ are included in the $20\leq \log N_{\rm H}<21$ bin). 
{ At $z=0$, the $N_{\rm H}$ function of U14, which is based on the 9-month {\sl Swift BAT}, AMSS and 
SXDS surveys, is in good agreement with that of \citet{burlon11}, which is  based on the three-year
{\sl Swift BAT} survey.}

 Also U14 generated a set of sophisticated template X-ray spectra of AGNs, which include (i) an underlying power-law continuum 
with a high energy cutoff at $E_{\rm c}=300$ keV, (ii) a reflection component, (iii) a scattering component from surrounding gas outside the 
torus, corresponding to 1\% with a torus opening solid angle is 2$\pi$, and an intrinsic absorption 
with Compton-scattering derived from Monte-Carlo simulations. For our purpose, we take the spectral energy
distribution (SED) templates in Fig. 7.  of U14 for $\log N_{\rm H}{\rm [cm^{-2}]}=20.5,21.5,22.5$, 
and 23.5, but ignored the luminosity and redshift dependence of the SED. In these template spectra, $\Gamma=1.94$ 
and $\Gamma=1.84$ are assumed for the primary power-law continuum for $\log N_{\rm H}{\rm [cm^{-2}]}<22$ and 
$>22$ AGNs respectively.  

 In this section, we refine our definitions. The symbol $L_{\rm X}$ represents an intrinsic
rest-frame 2-10 keV luminosity of the primary power-law, which does not include the reflection component. This is 
because including the reflection component double counts the intrinsic isotropic luminosity of the AGN. The reflection 
component is typically $\sim 15\%$ of the primary power-law in the 2-10 keV band. The symbol $d\Phi/d\log L_{\rm X}$
is the comoving number density of AGNs per dex in $L_{\rm X}$ for all Compton-thin (CTN) AGNs ($\log N_{\rm H}<24.0$).

 To feed realistic spectra into the maximum likelihood procedure, the $N$ function, Eq. \ref{eq:n_func}, should be 
a function of observable quantities for each sample. We define the ``observed'' 2-10 keV luminosity defined in Eq. \ref{eq:lobs}.

 In this case,  clearly distinguishing between observed and intrinsic X-ray luminosities, 
Eq. \ref{eq:n_func} can be rewritten as:
\begin{eqnarray}
&N(\log\;L_{{\rm X,obs}},z) d\,\log L_{{\rm X,obs}}dz=\nonumber \\
&\int_{20}^{24} \left(\frac{{\rm d}\,\Phi}{{\rm d}\log\,L_{\rm X}}d\,\log L_{\rm X}\right)\;d_{\rm A}(z)^2\;(1+z)^2  
  \frac{{\rm d}\chi}{{\rm d}z}(z)dz\nonumber \\
&\times f(L_{\rm X},z;N_{\rm H})A[S_{\rm X,obs}(L_{\rm X},N_{\rm H},z)]d \log N_{\rm H},
\label{eq:n_func_abs}
\end{eqnarray} 
where $S_{\rm X,obs}$ is now a function of $N_{\rm H}$ in addition to $L_{\rm X}$ and $z$. The integral is
over $\log N_{\rm H}$ and the intrinsic luminosity $L_{\rm X}$ is a unique function of $L_{\rm X,obs}$, $N_{\rm H}$,
and $z$. The ratio of $d\,\log L_{{\rm X,obs}}/d\,\log L_{\rm X}$ becomes non-unity only if the spectrum
changes with $L_{\rm X}$ for a given $N_{\rm H}$, which is the case for the {\sl Swift} BAT sample (see below). 
 
In each template spectrum, with an intrinsic 2-10 keV luminosity of the power-law continuum 
$L_{\rm X}$, we can calculate the observed luminosity at $2(1+z)-10(1+z)$ keV ($L_{\rm X,tmp}^{\rm obs}(N_{\rm H},z)$).
The true 2-10 keV flux from this object, under this template spectrum, is now:
\begin{equation}
S_{\rm X,obs}^{\rm tru}=L_{\rm X}/4\pi d_{\rm L}(z)^2\cdot \frac{L_{\rm X,tmp}^{\rm obs}(N_{\rm H},z)}{L_{\rm X}}.
\label{eq:sxobs_true}
\end{equation}
 
 The difference between $S_{\rm X,obs}^{\rm tru}$ to  $S_{\rm X,obs}^{\Gamma=1.7}$ (or the catalogued flux
with an ECF calculated with $\Gamma=1.7$ or our assumed spectrum for the ECF) is often neglected in the literature. 
In reality, it becomes important for highly absorbed AGNs ($\log N_{\rm H}\ga 23$) at low redshifts 
($z\lesssim 1$) in ``2-10 keV'' surveys, because the effective areas of 
the instruments used in generating our sample other than  {\sl Swift} BAT, drop rapidly as we go from 
2 keV to 10 keV. Thus the ECF increases rapidly as the spectrum gets harder. 
    
 An ideal approach would be to use separate response curve and countrate-survey area relation 
for each survey, as U14 did, since each survey use different instruments 
and/or different observational epochs, for which the calibration may vary.  However, a limitation of the software that 
we use for this work is that it can only accept one flux-limit -- solid angle curve, as the software is a legacy from 
a series of single-band soft X-ray luminosity work \citet{miyaji00a,miyaji01,hasinger05}.  
While this reduces the accuracy of the estimate of $S_{\rm x,obs}^{\rm tru}$, the calculation speed 
gained by this simplification has allowed us to explore analytical forms for various sub-sets 
and cases (see below).  We find that the relative ECFs between those for absorbed AGN spectra and for $\Gamma=1.7$ is not very
sensitive to which instrument we use. For example, the ratios of the ECFs for an extremely hard object 
($\Gamma=1.7$ with $\log N_{\rm NH}=23.5$ at z=0.2) to that for an unabsorbed $\Gamma=1.7$ spectrum are 2.0, 2.1, and 1.7 for 
the {\sl XMM-Newton} PN+2MOS (countrate at 2-8 kV), the {\sl Chandra} ACIS-I Cycle 8 (countrate at 2-7 keV) and 
the {\sl ASCA} GIS (CR at 2-10 keV) respectively. Here we use the response for ACIS-I for the conversion between
$S_{\rm x,obs}^{\rm tru}$ to  $S_{\rm x,obs}^{\Gamma=1.7}$. A caveat is that this causes some inaccuracy for 
the HBSS (CR is at 4.5-7.5 keV) and {\sl XMM-Newton} LH surveys (CR is at 2-4.5 keV). However,  the impact is 
relatively minor, since these latter surveys contain relatively small fraction of sources in the relevant flux range, i.e., the HBSS
is overwhelmed by the ASCA MSS+LSS and the LH sample by the COSMOS sample in their respective flux coverages. 

 For the {\sl Swift} BAT sample, the selections are made using the 15-55 keV range, which is very little affected by the 
absorption up to $\log N_{\rm H}=24$.  Despite the limitation of our software described above, 
the {\sl Swift} BAT sample can be treated separately because the flux regime covered by this sample 
does not overlap with any other. We have made the conversion between the 2-10 keV and the 15-55 keV bands
as follows.

 In the comparison between {\sl Swift} 14-194 keV and de-absorbed {\sl MAXI} 2-10 keV luminosities, \citet{ueda11} 
pointed out that the effective power-law index between these bands varies with luminosity from $\Gamma_{\rm l}=1.7$ to 
$\Gamma_{\rm u}=2.0$.  We have confirmed this by comparing the 14-55 keV luminosities from the \citep{ajello12} catalog 
with 4-10 keV luminosities in the latest {\sl MAXI} catalog of \citet{hiroi13}. To approximately represent the change, we use 
a smoothly varying luminosity-dependent effective spectral index
\begin{equation}
\Gamma_{\rm eff}=\Gamma_{\rm l}+0.5\;\mathrm{erf}[(\log L_{\rm X}-44.0)/2.0](\Gamma_{\rm u}-\Gamma_{\rm l})
\label{eq:ldepgamma}
\end{equation} 
for the conversion between the 2--10 keV and 15--55 keV fluxes, where $\mathrm{erf}$ is the error function. 
U14 hypothesize that the main cause of this slope change is the difference between type 1 and type 2 AGNs in the strength 
of component reflected by the torus (i.e., the torus tend to be thicker in type 2 AGNs) 
in combination with the relation that the type 2 fraction decreases with luminosity.  
Because of the definition of $L_{\rm X}$ in this section, the 2-10 keV flux is calculated based on $(1+0.15)L_{\rm X}$
following the  assumption that 15\% of intrinsic 2-10 keV luminosity is added by the reflection component. 
A small K-correction has been made based on $\Gamma_{\rm eff}$. In order to use our flux limit -- survey area curve, 
we have back-converted from the  15-55 keV flux to observed flux $S_{\rm X,obs}^{\Gamma=1.7}$ using our reference 
slope $\Gamma=1.7$. 

As a technical note, we comment that U14 chose to use the count rate as an independent variable to the N-function 
instead of $\log\; L_{\rm X, obs}$.

\subsection{The $N_{\rm obs}/N_{\rm mdl}$ method for the Binned XLF}
\label{sec:nobs_nmdl}

 A less biased way of presenting the binned XLF than the classical $\sum 1/V_{\rm a}$ 
\citep{avni80} is the $N_{\rm obs}/N_{\rm mdl}$ estimator described in \citet{miyaji01} 
\citep[see also][]{lafranca97}. 
The basic procedure is:
\begin{enumerate}
\item Divide the combined sample into several redshift shells. 
  For each redshift shell, fit the AGN XLF with a smooth analytical
  function obtained with the maximum-likelihood fit as described above. 
\item For each redshift shell, count the number of objects that fall 
  into each luminosity bin to obtain the observed number of objects ($N_{\rm obs}$).
\item For each luminosity bin, evaluate the analytical
  fit at the central luminosity/redshift 
 ($d\Phi^{\rm mdl}/d{\rm log} L_{\rm X}$).
\item Calculate the predicted number of AGNs in the bin ($N_{\rm mdl}$).
\item The final result is: 
  \begin{equation}
  \frac{d\Phi}{d\,{\rm log\,}L_{\rm X}}=\frac{d\Phi^{\rm mdl}}{d\;{\rm log\,}L_{\rm X}} \cdot
  \;N_{\rm obs}/N_{\rm mdl}. 
  \end{equation}
\end{enumerate}

 We note that, in the case that the XLF model is constant within the bin, the $N_{\rm obs}/N_{\rm mdl}$
estimator is mathematically equivalent to that proposed by \citet{page00}.  
In the case of taking absorption into account with the method described above in
Sect. \ref{sec:xlf_abs}, and binning is by $\log L_{\rm X}$ defined there, deriving $N_{\rm obs}$ is
not straightforward, since we cannot determine  $\log L_{\rm X}$ individually from single-band 
fluxes. The $N_{\rm H}$ values derived from hardness ratios are subject to large errors and 
not necessarily available for all objects in the sample.  On the other hand,
it is straightforward to calculate from the model probability distribution of $N_{\rm H}$ for given
redshift and observed flux: $p(z, S_{\rm x,obs};\log N_{\rm H})$, which can be normalized as: 
\begin{equation}
\int_{20}^{24}p(z, S_{\rm x,obs};\log N_{\rm H})d\log N_{\rm H}=1.
\end{equation}

 This is different from the $N_{\rm H}$ function $f(L_{\rm X},z;\log N_{\rm H})$,
which is the distribution of $N_{\rm H}$ for given redshift and intrinsic luminosity. For our 
purpose, for the $i$-th object with a given $(z_i, S_{{\rm x,obs},i})$, we use the probability 
distribution of the intrinsic luminosity: 
\begin{eqnarray}
p_i(\log L_{\rm X}[(z_i, S_{{\rm x,obs},i},N_{\rm H})])d\log L_{\rm X}=\nonumber\\
p(z_i, S_{{\rm x,obs},i};\log N_{\rm H})d\log N_{\rm H}.
\end{eqnarray}

If we would like to calculate $N_{\rm obs}$ in the redshift-intrinsic luminosity bin $(z_1,z_2;\log L_1,\log L_2)$:
\begin{equation}
N_{\rm obs}=\sum_i\int_{\rm \log L_1}^{\rm log L_2}p_i(\log L_{\rm X})d\log L_{\rm X},
\end{equation}
where $i$ is over objects with redshifts between $z_1$ and $z_2$. In practice, we calculate $N_{\rm obs}$ as
weighted sums over $\log N_{\rm H}=20.5,21.5,22.5$ and 23.5 templates. 
Similarly, we include the PDFs of photometric redshifts into $N_{\rm obs}$ estimations by adding
the portion of the PDF of each objects that falls into the redshift bin, as has been made also 
by \citet{aird10}. 

 One limitation is that the errors assuming Poisson statistics are not valid if $N_{\rm obs}$ is a
sum of weights, unlike the cases of our previous works \citep[][U14]{miyaji01,hasinger05}, in 
which $N_{\rm obs}$ were always integers. However, we use the approximate 1$\sigma$ Poisson errors 
from the \citet{gehrels86} Eqs. (7) and (12) ($S=1$) for the upper and lower 1$\sigma$ errors respectively.  
Thus, the errors are only approximate.  Also the errors in different bins are not independent of 
one another.   

\section{Global XLF Expression}
\label{sec:xlf_global}

\subsection{The Luminosity-dependent Density Evolution Model}

 For the analytical expression of the XLF at z=0, we use 
the {\em smoothed two power-law} formula. 
\begin{eqnarray}
\frac{{\rm d}\;\Phi\,(L_{\rm X},0)}{{\rm d\;log}\;L_{\rm X}}&=&
A_{44}^{z=0}\frac{
\left(\frac{10^{44}}{{L_{\rm x,*}}}\right)^{{\gamma_1}}
         +\left(\frac{10^{44}}{{L_{\rm x,*}}}\right)^{{\gamma_2}}
}{
\left(\frac{L_{\rm X}}{{L_{\rm x,*}}}\right)^{{\gamma_1}}
         +\left(\frac{L_{\rm X}}{{L_{\rm x,*}}}\right)^{{\gamma_2}}
}\nonumber \\
&\equiv&\frac{A_*^{z=0}}{
  \left(\frac{L_{\rm X}}{{L_{\rm x,*}}}\right)^{{\gamma_1}}
  +\left(\frac{L_{\rm X}}{{L_{\rm x,*}}}\right)^{{\gamma_2}}.
}
\label{eq:2po}
\end{eqnarray}
The normalization $A_{\rm 44}^{z=0}$ is the XLF value at $(\log L_{\rm X},z)=(44,0)$. There is also a convention
to use $A_*$ (2nd line in Eq. \ref{eq:2po}) as a normalization parameter.  We primarily use $A_{\rm 44}^{z=0}$, 
because $A_*$ is strongly coupled with the break luminosity $L_{X,*}$, and therefore it is not possible 
to estimate its error accurately with our maximum-likelihood fitting procedure (see below).   

 Following the most successful analytical form of the evolution in the literature, we 
use the luminosity-dependent density evolution: 

\begin{equation}
\frac{{\rm d}\;\Phi\,(L_{\rm X},z)}{{\rm d\;log}\;L_{\rm X}}
  = \frac{{\rm d}\;\Phi\,(L_{\rm X},0)}
    {{\rm d\;log}\;L_{\rm X}}\cdot e_{\rm d}(z, L_{\rm X}),
\label{eq:ldde0}
\end{equation}

where $e_{\rm d}(z, L_{\rm X})$ is the density evolution factor normalized to z=0.
We describe it with a three-segment power-law model: 

\begin{equation}
e_{\rm d}(z,L_{\rm X}) = \left\{
\begin{array}{ll}
 (1+z)^{p_1} & (z<z_{\rm b1})\\
 e_{\rm d}(z_{\rm b1},L_{\rm X})\cdot\left(\frac{1+z}{1+z_{\rm b1}}\right)^{p_2} & (z_{\rm b1}\leq z < z_{\rm b2})\\
 e_{\rm d}(z_{\rm b2},L_{\rm X})\cdot\left(\frac{1+z}{1+z_{\rm b2}}\right)^{p_3} & (z \geq z_{\rm b2}).
\end{array}
\right. 
\label{eq:3po_ev}
\end{equation}

 As in the previous works, the first break positions $z_{\rm p1}$ may depend on luminosity at the low luminosity 
regime:  
\begin{equation}
z_{\rm b1}(L_{\rm X}) = \left\{ 
	\begin{array}{ll}
	z_{\rm b,0}(L_{\rm X}/L_{\rm x,b})^\alpha & 
	(L_{\rm X} \leq L_{\rm x,b}) \\ 
        z_{\rm b,0} & (L_{\rm X}> L_{\rm x,b})\\
	\end{array}
       \right. .
\label{eq:ldde2}
\end{equation}

We also include the luminosity dependence of the evolution indices $p1$ and $p2$: 
\begin{eqnarray}
p_1 (L_{\rm X}) &=& p_{1,44}+ \beta_1\;(\log L_{\rm X}-44)\\
p_2 (L_{\rm X}) &=& p_{2,44}+ \beta_2\;(\log L_{\rm X}-44).
\end{eqnarray}

\subsection{Global LDDE Results}
\label{sec:ldde_res}

 The best-fit parameters of our LDDE model and 1$\sigma$ errors, corresponding to  
$\Delta \mathcal{L}=1$, are shown in Table \ref{tab:best}. 
For the global expression, the fits using the likelihood function Eq. \ref{eq:ml2} give much 
more stable error search results than those using Eq. \ref{eq:ml1}. 
As stated above, minimizing $\mathcal{L}$ in Eq. \ref{eq:ml2} cannot determine the normalization.
Thus the $A_{44}^{z=0}$ value and its error are determined by scaling it such that it gives the total
number of objects ($N_{\rm tot}$) for best-fit values of other parameters. The fractional error of 
$1/\sqrt{N_{\rm tot}}$ is used for its error. Thus the errors of $A_{44}^{z=0}$ in Table \ref{tab:best} 
do not contain correlation of errors with other parameters.       
The traditional normalization $A_*^{z=0}$ (see Eq. \ref{eq:2po}) is also shown without an error 
for reference. 

 The parameters for the fit in which the realistic AGN spectra, $N_{\rm H}$ function
(Sect. \ref{sec:xlf_abs}), and the PDFs of photometric redshifts (Sect. \ref{sec:pzpdf}) are fully
taken into account are listed under ``Full'' in Table \ref{tab:best}. For our final results, we
show the case where $z_{\rm b1,44}$ is fixed to 1.1 and $z_{\rm b1,44}=2.7$, since 
fitting with all parameters free causes failures in the minimization processes. The values of
these fixed parameters have been decided based on the results of the fits in 
luminosity-class divided samples, which is explained later in Sect. \ref{sec:evol}. Since the fits in each 
luminosity class also use information on the global expression, thus the process has been iterative, i.e.,
we decide on fixed parameters of the global expression referring to the results of the luminosity-class
divided sample based on the global expression in the previous step.

\begin{deluxetable}{lccc}
\tabletypesize{\footnotesize}
\tablecaption{Best--fit LDDE parameters for Global Expressions \label{tab:best}}
\tablewidth{0pt}
\tablehead{
 \colhead{Parameter\tablenotemark{a}} & \colhead{Full} & \colhead{No pz-PDZ} & \colhead{$\Gamma=1.8$}
}
\startdata
\cutinhead{z=0 XLF Parameters\tablenotemark{b}}
$A_{44}^{z=0}$\tablenotemark{c} & $(9.3\pm 0.2)\times 10^{-7}$ & $(8.7\pm 0.2)\times 10^{-7}$ & $(9.7\pm 0.2)\times 10^{-7}$\\
$A_*^{z=0}$\tablenotemark{d}   &  $1.56 \times 10^{-6}$ & $1.34 \times 10^{-6}$  &  $1.60 \times 10^{-6}$  \\
$\log L_{\rm x,*}$        & $44.04\pm 0.08$ &  $44.06\pm 0.06$ & $44.05\pm 0.06$\\
$\gamma_1$                & $1.17\pm 0.05$  &  $1.17\pm 0.05$  & $1.09\pm 0.04$ \\
$\gamma_2$                & $2.80^{+0.16}_{-0.10}$  &  $2.90^{+0.08}_{-0.12}$ & $2.76\pm 0.08$\\
\cutinhead{Evolution Parameters\tablenotemark{b}}
$p_{1,44}$                 & $5.29\pm 0.11$  & $5.35\pm 0.11$ & $4.86\pm 0.11$ \\
$z_{\rm b1,44}$           & $1.1*$          & $1.1*$  & $1.1*$ \\
$p_{2,44}$                 & $-0.35\pm 0.14$ & $0.02\pm 0.13$ & $0.02\pm 0.14$\\
$z_{\rm b2}$              & $2.7*$          & $2.7*$  & $2.7*$ \\
$p3_{44}$                 & $-5.6*$         & $-5.6*$ & $-5.6*$ \\
$\alpha$                  & $0.18\pm .03$   & $0.16^{+.04}_{-.01}$ & $0.11^{+.01}_{-.02}$ \\
${\rm log } L_{\rm x,b}$  & $44.5*$         & $44.5*$   & $44.5*$\\
$\beta_1$                 & $1.2^{+0.3}_{-0.2}$    & $1.2^{+0.2}_{-0.3}$ & $1.6\pm 0.2$\\
$\beta_2$                 & $1.5^{+0.5}_{-0.3}$    & $2.0^{+0.2}_{-0.4}$ & $1.7\pm 0.2$\\
\enddata
\tablenotetext{a}{Parameters that have been fixed during the fit are labeled
  by '(*)'. Units - $A$ (with any sub and superscripts): $h_{70}^3\;{\rm Mpc^{-3}dex^{-1}}$,\,\,
  $L$ (with any sub and superscripts): $h_{70}^{-2}{\rm erg\;s^{-1}}$.}
\tablenotetext{b}{The 68\% confidence range for one parameter ($\Delta \mathcal{L}<1$) with correlations
  among parameter errors, except for $A_{44}^{z=0}$.}
\tablenotetext{c}{The 1$\sigma$ errors of $A_{44}^{z=0}$ is for all other parameters fixed.} 
\tablenotetext{d}{The traditional normalization defined as the twice of the XLF value
at the break luminosity at $z=0$ ($L_{\rm x,*}$).}
\end{deluxetable}



\section{XLFs in redshift shells and luminosity classes}
\label{sec:zshell_llclass}

\subsubsection{Adaptive binning}

 In this section, we present the binned XLF, as well as analytical formula of the XLF divided into 
redshift shells. We also present the number density curve divided in luminosity classes. The binning 
has been decided on a adaptive-basis, so that the sizes of error bars and bin-sizes are optimal
for the plots to be visible and informative. 

 First, we divide the sample into 11 redshift shells in the same way as those used by
U14, except that the two highest redshift shells by U14 ($z>3.0$) are combined into a single one
for obtaining a sufficient number of objects for a separate two power-law fit.
For each redshift shell, we construct luminosity bins such that each contains 20 objects by default, 
with minimum and maximum bin sizes of  $\Delta \log L_{\rm X}=0.125$ and 0.5 respectively. 
In case the most luminous bin contains less than 20 objects, the border between the 
last two bins is adjusted to contain approximately equal numbers of objects 
within the bin size constraints. 

 For the luminosity division, the range $42.0\leq \log L_{\rm X}\leq 46.0$ is divided into
four $\Delta \log L_{\rm X}=1.0$ classes. Each luminosity class is divided into redshift bins such 
that each bin has 20 objects (using the best fit photometric redshifts for those without spectroscopic
redshift) with forced minimum and maximum sizes of $\Delta \ln(1+z)=0.07$ and 0.5 respectively. 
If the highest redshift bin has less than 20 objects, the border
between the highest two redshift bins are adjusted in the same way as the case of luminosity 
binning. { Due to the small number of objects, the default number of objects per redshift bin for the   
$45.0\leq \log L_{\rm X}\leq 46.0$ class has been set to 7, with the same minimum and maximum bin size 
constraints.}

\subsection{Analytical form for each redshift shell}
 
 For the analytical expression of the XLF in each redshift shell, we use 
the {\em smoothed two power law} formula. 
Because the redshift shells have finite widths, the fit results depend on the evolution of the 
SXLF within them: 
\begin{eqnarray}
\frac{{\rm d}\;\Phi\,(L_{\rm X},z) }{{\rm d\;log}\;L_{\rm X}}&=&
A_{44}^{z}\frac{
\left(\frac{10^{44}}{{L_{\rm x,*}}}\right)^{{\gamma_1}}
         +\left(\frac{10^{44}}{{L_{\rm x,*}}}\right)^{{\gamma_2}}
}{
\left(\frac{L_{\rm X}}{{L_{\rm x,*}}}\right)^{{\gamma_1}}
         +\left(\frac{L_{\rm X}}{{L_{\rm x,*}}}\right)^{{\gamma_2}}
}\nonumber\\
 &\times& e_{\rm d}(z,L_{\rm X}),
\label{eq:2poz}
\end{eqnarray}
where $e_{\rm d}(z,L_{\rm X})$ is the density evolution factor.
While the final results are insensitive to the detailed behavior of  
$e_{\rm d}(z,L_{\rm X})$ within the shell at most locations in the 
$(L_{\rm X},z)$ space, we have taken our best-estimate by using the
luminosity-dependent density evolution (LDDE) model derived above in 
Sect. \ref{sec:xlf_global}. The luminosity range of the fit is from 
$\log \,L_{\rm X,obs}$=41.5 to 46.0.

The best fit parameters of Eq.~\ref{eq:2poz} for each redshift shell are 
shown in Table~\ref{tab:zshell2po}. The normalization is 
defined by:
\begin{equation}
A_{\rm 44}^{z=z_{\rm c}}=\frac{{\rm d}\;\Phi\,(L_{\rm X}=10^{44}\;{\rm erg\;s^{-1}} ,z=z_{\rm c}) }
{{\rm d\;log}\;L_{\rm X}},
\end{equation}   
where $z_{\rm c}$ is the central redshift of the shell, which is defined by
$z_{\rm c}+1=\sqrt{(1+z_{\rm min})(1+z_{\rm max})}$, where $z_{\rm min}$ and
$z_{\rm max}$ are the minimum and maximum redshifts of the bin. One sigma parameter errors are calculated in the same way as 
in Sect. \ref{sec:ldde_res}. In this section, we have been able to obtain 1$\sigma$ errors of most 
parameters with Eq. \ref{eq:ml1}, and unlike the case of Sect. \ref{sec:ldde_res}, the errors
of $A_{\rm 44}^{z=z_{\rm c}}$ contain the correlations among parameter errors.  
We also show the traditional normalization $A_*^{z=z_{\rm c}}$ without errors.  

\begin{deluxetable}{ccccccc}
\tabletypesize{\footnotesize}
\tablecaption{Smoothed 2PL parameters for each redshift shell\label{tab:zshell2po}}
\tablewidth{0pt}
\tablehead{
\colhead{$z$-range} &\colhead{$z_{\rm c}$} & \colhead{$A_{44}^{z=z_{\rm c}}$ \tablenotemark{a}\tablenotemark{b}} 
&\colhead{$A_{\rm *}^{z=z_{\rm c}}$ \tablenotemark{a}\tablenotemark{b}} & \colhead{$\log L_*^{z=z_{\rm c}}$ \tablenotemark{b}} 
&\colhead{$\gamma_1$\tablenotemark{a}} & \colhead{$\gamma_2$\tablenotemark{a}}
}
\startdata
 0.015 - 0.200 & 0.104 & $1.20^{+0.14}_{-0.13}\times 10^{-6}$ & $8.97 \times 10^{-6}$ & $ 43.70^{+0.18}_{-0.19}$ & $ 0.90^{+0.14}_{-0.17}$ & $ 2.53^{+0.23}_{-0.19}$\\
 0.200 - 0.400 & 0.296 & $4.79^{+0.65}_{-0.61}\times 10^{-6}$ & $3.91 \times 10^{-6}$ & $ 44.22^{+0.21}_{-0.21}$ & $ 1.05^{+0.08}_{-0.10}$ & $ 2.90^{+0.37}_{-0.27}$\\
 0.400 - 0.600 & 0.497 & $7.64^{+1.53}_{-1.38}\times 10^{-6}$ & $1.67 \times 10^{-5}$ & $ 43.98^{+0.20}_{-0.23}$ & $ 0.83^{+0.11}_{-0.14}$ & $ 2.70^{+0.48}_{-0.32}$\\
 0.600 - 0.800 & 0.697 & $1.14^{+0.17}_{-0.14}\times 10^{-5}$ & $2.36 \times 10^{-4}$ & $ 43.31^{+0.29}_{-0.24}$ & $ 0.41^{+0.24}_{-0.30}$ & $ 1.86^{+0.17}_{-0.13}$\\
 0.800 - 1.000 & 0.897 & $3.12^{+0.37}_{-0.35}\times 10^{-5}$ & $8.91 \times 10^{-5}$ & $ 43.90^{+0.13}_{-0.13}$ & $ 0.40^{+0.13}_{-0.15}$ & $ 2.51^{+0.27}_{-0.22}$\\
 1.000 - 1.200 & 1.098 & $3.47^{+0.35}_{-0.33}\times 10^{-5}$ & $1.06 \times 10^{-4}$ & $ 43.85^{+0.14}_{-0.14}$ & $ 0.19^{+0.16}_{-0.19}$ & $ 2.06^{+0.18}_{-0.15}$\\
 1.200 - 1.600 & 1.392 & $4.42^{+0.31}_{-0.29}\times 10^{-5}$ & $3.04 \times 10^{-5}$ & $ 44.43^{+0.09}_{-0.10}$ & $ 0.47^{+0.07}_{-0.08}$ & $ 2.82^{+0.32}_{-0.27}$\\
 1.600 - 2.000 & 1.793 & $3.68^{+0.32}_{-0.29}\times 10^{-5}$ & $2.07 \times 10^{-5}$ & $ 44.57^{+0.10}_{-0.11}$ & $ 0.48^{+0.09}_{-0.10}$ & $ 2.73^{+0.38}_{-0.32}$\\
 2.000 - 2.400 & 2.194 & $4.27^{+0.50}_{-0.43}\times 10^{-5}$ & $3.28 \times 10^{-5}$ & $ 44.50^{+0.12}_{-0.13}$ & $ 0.27^{+0.14}_{-0.17}$ & $ 2.84^{+0.52}_{-0.41}$\\
 2.400 - 3.000 & 2.688 & $3.28^{+0.48}_{-0.33}\times 10^{-5}$ & $1.71 \times 10^{-5}$ & $ 44.63^{+0.09}_{-0.17}$ & $ 0.46^{+0.12}_{-0.17}$ & $ 3.12^{+\infty}_{-0.60}$\\
 3.000 - 5.800 & 4.215 & $4.37^{+0.90}_{-0.62}\times 10^{-6}$ & $1.08 \times 10^{-6}$ & $ 44.94^{+0.20}_{-0.28}$ & $ 0.65^{+0.19}_{-0.25}$ & $ 4.38^{+\infty}_{-1.69}$\\
\enddata
\tablenotetext{a}{Errors show the 68\% confidence range for one parameter ($\Delta \mathcal{L}<1$).}
\tablenotetext{b}{Units -- $A$ (with any sub/upper scripts): [$h_{70}^3\;{\rm Mpc^{-3}dex^{-1}}$],\,\,
$L$ (with any sub/upper scripts): $[\;h_{70}^{-2}{\rm erg\;s^{-1}}]$.}
\end{deluxetable}

\begin{figure*}
\begin{center}
\resizebox{!}{!}{
  \includegraphics[width=\hsize]{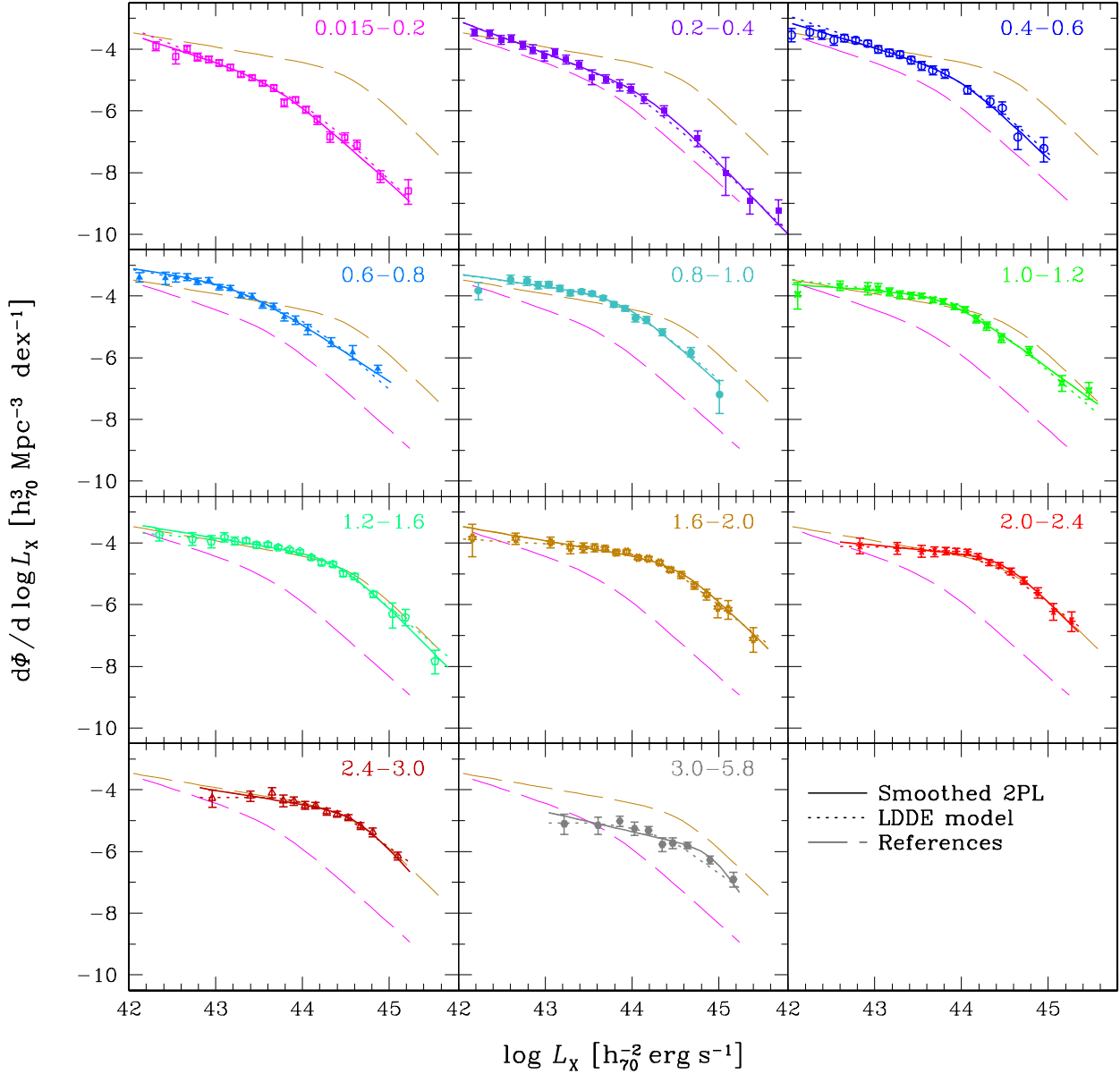}
}
\end{center}

\caption{The $N_{\rm obs}/N_{\rm mdl}$ estimated intrinsic 2-10 keV X--ray luminosity function 
 is plotted in each redshift shell as labeled. The 68\% approximate Poisson error bars are shown 
 in each data point. The solid line in each panel shows the 
 redshift divided smoothed 2PL model at the central redshift ($z_{\rm c}$). The best-fit 
 LDDE model is shown with dotted lines.  The best--fit smoothed two power law models for 
 the lowest redshift shell ($z_{\rm c}=0.093$) and at $z_{\rm c}=2.214$ are shown with 
 dashed lines in all other panels for reference. (Color version of this figure is available 
 in the electronic version.)}
\label{fig:exlf}
\end{figure*}

 Table~\ref{tab:exlf_zshell} tabulates the binned $N_{\rm obs}/N_{\rm mdl}$-estimated XLF for the full treatment case,
along with the observed number of AGNs ($N_{\rm obs}$) and the final estimated values XLF value at the center of each bin.  
The full XLFs in the 11 redshift shells are plotted in Fig.~\ref{fig:exlf} in separate panels.
The best-fit smoothed two power-law model (2PL) and the best-fit LDDE model evaluated at $z_{\rm c}$
are overplotted to assess the goodness of the models. The best--fit two power law models at 
$z_{\rm c}=0.1$ (lowest redshift bin) and at $z_{\rm c}=1.8$, where the $A_{\rm 44}$ 
is near the peak, are also shown in all panels for reference. 

\begin{deluxetable}{cccccc}
\tabletypesize{\footnotesize}
\tablecaption{Full binned XLF divided in redshift shells\label{tab:exlf_zshell}}
\tablewidth{0pt}
\tablehead{
  \colhead{$z$-range} &\colhead{$z_{\rm c}$} &\colhead{$\log L_{\rm X}$-range\tablenotemark{a}} & 
  \colhead{$\log L_{\rm X,c}$\tablenotemark{a}} & \colhead{$N_{\rm obj}$} & \colhead{$\frac{d\Phi}{d\log\, L_{\rm X}}$\tablenotemark{a,b}}
}
\startdata
0.015 - 0.200 & 0.104 & 42.13 - 42.48 & 42.31 &  11.4 & $(1.24\pm 0.24)\times 10^{-4}$ \\
0.015 - 0.200 & 0.104 & 42.48 - 42.61 & 42.54 &   5.2 & $(5.85\pm 1.76)\times 10^{-5}$ \\
0.015 - 0.200 & 0.104 & 42.61 - 42.73 & 42.67 &  15.9 & $(1.02\pm 0.23)\times 10^{-4}$ \\
0.015 - 0.200 & 0.104 & 42.73 - 42.86 & 42.79 &  16.1 & $(5.58\pm 1.28)\times 10^{-5}$ \\
0.015 - 0.200 & 0.104 & 42.86 - 42.98 & 42.92 &  20.0 & $(4.68\pm 0.89)\times 10^{-5}$ \\
0.015 - 0.200 & 0.104 & 42.98 - 43.11 & 43.04 &  26.1 & $(3.61\pm 0.65)\times 10^{-5}$ \\
0.015 - 0.200 & 0.104 & 43.11 - 43.23 & 43.17 &  26.4 & $(2.54\pm 0.46)\times 10^{-5}$ \\
0.015 - 0.200 & 0.104 & 43.23 - 43.36 & 43.29 &  26.6 & $(1.54\pm 0.26)\times 10^{-5}$ \\
0.015 - 0.200 & 0.104 & 43.36 - 43.48 & 43.42 &  28.7 & $(1.20\pm 0.20)\times 10^{-5}$ \\
0.015 - 0.200 & 0.104 & 43.48 - 43.61 & 43.54 &  29.6 & $(7.81\pm 1.33)\times 10^{-6}$ \\
0.015 - 0.200 & 0.104 & 43.61 - 43.73 & 43.67 &  28.2 & $(5.55\pm 1.00)\times 10^{-6}$ \\
0.015 - 0.200 & 0.104 & 43.73 - 43.86 & 43.79 &  14.8 & $(1.85\pm 0.46)\times 10^{-6}$ \\
0.015 - 0.200 & 0.104 & 43.86 - 43.98 & 43.92 &  24.4 & $(2.25\pm 0.45)\times 10^{-6}$ \\
0.015 - 0.200 & 0.104 & 43.98 - 44.11 & 44.04 &  18.9 & $(1.08\pm 0.25)\times 10^{-6}$ \\
0.015 - 0.200 & 0.104 & 44.11 - 44.23 & 44.17 &  12.5 & $(5.18\pm 1.45)\times 10^{-7}$ \\
0.015 - 0.200 & 0.104 & 44.23 - 44.42 & 44.32 &   9.2 & $(1.44\pm 0.48)\times 10^{-7}$ \\
0.015 - 0.200 & 0.104 & 44.42 - 44.55 & 44.49 &  10.0 & $(1.34\pm 0.43)\times 10^{-7}$ \\
0.015 - 0.200 & 0.104 & 44.55 - 44.70 & 44.63 &  11.0 & $(7.88\pm 2.36)\times 10^{-8}$ \\
0.015 - 0.200 & 0.104 & 44.70 - 45.10 & 44.90 &   7.0 & $(7.51\pm 2.85)\times 10^{-9}$ \\
0.015 - 0.200 & 0.104 & 45.10 - 45.35 & 45.22 &   2.0 & $(2.53\pm 1.77)\times 10^{-9}$ \\
\\                                                     
0.200 - 0.400 & 0.296 & 41.89 - 42.07 & 41.98 &   5.9 & $(2.48\pm 0.62)\times 10^{-4}$ \\
0.200 - 0.400 & 0.296 & 42.07 - 42.30 & 42.18 &  18.4 & $(3.50\pm 0.53)\times 10^{-4}$ \\
0.200 - 0.400 & 0.296 & 42.30 - 42.42 & 42.36 &  13.3 & $(3.11\pm 0.53)\times 10^{-4}$ \\
0.200 - 0.400 & 0.296 & 42.42 - 42.55 & 42.49 &  12.7 & $(2.03\pm 0.35)\times 10^{-4}$ \\
0.200 - 0.400 & 0.296 & 42.55 - 42.67 & 42.61 &  16.8 & $(2.21\pm 0.33)\times 10^{-4}$ \\
0.200 - 0.400 & 0.296 & 42.67 - 42.80 & 42.74 &  14.3 & $(1.36\pm 0.20)\times 10^{-4}$ \\
0.200 - 0.400 & 0.296 & 42.80 - 42.92 & 42.86 &  12.2 & $(9.95\pm 1.59)\times 10^{-5}$ \\
0.200 - 0.400 & 0.296 & 42.92 - 43.05 & 42.99 &  10.1 & $(6.07\pm 1.03)\times 10^{-5}$ \\
0.200 - 0.400 & 0.296 & 43.05 - 43.17 & 43.11 &  15.0 & $(7.75\pm 1.08)\times 10^{-5}$ \\
0.200 - 0.400 & 0.296 & 43.17 - 43.30 & 43.24 &  11.7 & $(4.67\pm 0.75)\times 10^{-5}$ \\
0.200 - 0.400 & 0.296 & 43.30 - 43.48 & 43.39 &  13.5 & $(3.17\pm 0.48)\times 10^{-5}$ \\
0.200 - 0.400 & 0.296 & 43.48 - 43.60 & 43.54 &   4.9 & $(1.25\pm 0.33)\times 10^{-5}$ \\
0.200 - 0.400 & 0.296 & 43.60 - 43.80 & 43.70 &  12.1 & $(1.08\pm 0.18)\times 10^{-5}$ \\
0.200 - 0.400 & 0.296 & 43.80 - 43.92 & 43.86 &   8.4 & $(6.79\pm 1.56)\times 10^{-6}$ \\
0.200 - 0.400 & 0.296 & 43.92 - 44.05 & 43.99 &  11.4 & $(5.31\pm 1.12)\times 10^{-6}$ \\
0.200 - 0.400 & 0.296 & 44.05 - 44.23 & 44.14 &  11.2 & $(2.49\pm 0.52)\times 10^{-6}$ \\
0.200 - 0.400 & 0.296 & 44.23 - 44.51 & 44.37 &   9.8 & $(1.01\pm 0.24)\times 10^{-6}$ \\
0.200 - 0.400 & 0.296 & 44.51 - 45.01 & 44.76 &   4.3 & $(1.28\pm 0.51)\times 10^{-7}$ \\
0.200 - 0.400 & 0.296 & 45.01 - 45.16 & 45.08 &   1.1 & $(9.41\pm 8.85)\times 10^{-9}$ \\
0.200 - 0.400 & 0.296 & 45.16 - 45.56 & 45.36 &   2.0 & $(1.28\pm 0.90)\times 10^{-9}$ \\
0.200 - 0.400 & 0.296 & 45.56 - 45.81 & 45.69 &   2.0 & $(5.77\pm 4.04)\times 10^{-10}$ \\
\\                                                     
0.400 - 0.600 & 0.497 & 41.90 - 42.19 & 42.04 &   5.9 & $(2.92\pm 0.70)\times 10^{-4}$ \\
0.400 - 0.600 & 0.497 & 42.19 - 42.31 & 42.25 &   6.3 & $(3.58\pm 0.86)\times 10^{-4}$ \\
0.400 - 0.600 & 0.497 & 42.31 - 42.46 & 42.39 &  10.5 & $(2.96\pm 0.56)\times 10^{-4}$ \\
0.400 - 0.600 & 0.497 & 42.46 - 42.59 & 42.53 &   9.8 & $(1.96\pm 0.39)\times 10^{-4}$ \\
0.400 - 0.600 & 0.497 & 42.59 - 42.71 & 42.65 &  15.3 & $(2.29\pm 0.34)\times 10^{-4}$ \\
0.400 - 0.600 & 0.497 & 42.71 - 42.85 & 42.78 &  22.1 & $(1.95\pm 0.23)\times 10^{-4}$ \\
0.400 - 0.600 & 0.497 & 42.85 - 42.98 & 42.92 &  22.7 & $(1.53\pm 0.18)\times 10^{-4}$ \\
0.400 - 0.600 & 0.497 & 42.98 - 43.10 & 43.04 &  17.0 & $(9.81\pm 1.28)\times 10^{-5}$ \\
0.400 - 0.600 & 0.497 & 43.10 - 43.23 & 43.17 &  17.3 & $(7.55\pm 0.98)\times 10^{-5}$ \\
0.400 - 0.600 & 0.497 & 43.23 - 43.35 & 43.29 &  18.0 & $(6.69\pm 0.87)\times 10^{-5}$ \\
0.400 - 0.600 & 0.497 & 43.35 - 43.48 & 43.42 &  16.7 & $(4.45\pm 0.58)\times 10^{-5}$ \\
0.400 - 0.600 & 0.497 & 43.48 - 43.60 & 43.54 &  12.7 & $(2.85\pm 0.43)\times 10^{-5}$ \\
0.400 - 0.600 & 0.497 & 43.60 - 43.73 & 43.67 &  11.8 & $(2.08\pm 0.33)\times 10^{-5}$ \\
0.400 - 0.600 & 0.497 & 43.73 - 43.88 & 43.81 &  11.8 & $(1.60\pm 0.27)\times 10^{-5}$ \\
0.400 - 0.600 & 0.497 & 43.88 - 44.25 & 44.07 &  13.2 & $(4.79\pm 0.77)\times 10^{-6}$ \\
0.400 - 0.600 & 0.497 & 44.25 - 44.41 & 44.33 &   7.9 & $(2.02\pm 0.55)\times 10^{-6}$ \\
0.400 - 0.600 & 0.497 & 44.41 - 44.53 & 44.47 &   6.5 & $(1.25\pm 0.39)\times 10^{-6}$ \\
0.400 - 0.600 & 0.497 & 44.53 - 44.77 & 44.65 &   2.3 & $(1.42\pm 0.54)\times 10^{-7}$ \\
0.400 - 0.600 & 0.497 & 44.77 - 45.13 & 44.95 &   2.1 & $(6.14\pm 3.50)\times 10^{-8}$ \\
\\                                                     
0.600 - 0.800 & 0.697 & 41.88 - 42.36 & 42.12 &  10.9 & $(4.06\pm 0.69)\times 10^{-4}$ \\
0.600 - 0.800 & 0.697 & 42.36 - 42.48 & 42.42 &   6.3 & $(3.83\pm 0.84)\times 10^{-4}$ \\
0.600 - 0.800 & 0.697 & 42.48 - 42.61 & 42.54 &  11.8 & $(4.05\pm 0.69)\times 10^{-4}$ \\
0.600 - 0.800 & 0.697 & 42.61 - 42.73 & 42.67 &  18.4 & $(4.20\pm 0.55)\times 10^{-4}$ \\
0.600 - 0.800 & 0.697 & 42.73 - 42.86 & 42.79 &  20.9 & $(2.99\pm 0.39)\times 10^{-4}$ \\
0.600 - 0.800 & 0.697 & 42.86 - 42.98 & 42.92 &  29.1 & $(3.26\pm 0.33)\times 10^{-4}$ \\
0.600 - 0.800 & 0.697 & 42.98 - 43.11 & 43.04 &  29.0 & $(1.96\pm 0.20)\times 10^{-4}$ \\
0.600 - 0.800 & 0.697 & 43.11 - 43.23 & 43.17 &  36.7 & $(1.83\pm 0.17)\times 10^{-4}$ \\
0.600 - 0.800 & 0.697 & 43.23 - 43.36 & 43.29 &  30.1 & $(1.08\pm 0.11)\times 10^{-4}$ \\
0.600 - 0.800 & 0.697 & 43.36 - 43.48 & 43.42 &  30.5 & $(9.83\pm 0.95)\times 10^{-5}$ \\
0.600 - 0.800 & 0.697 & 43.48 - 43.61 & 43.54 &  21.1 & $(5.09\pm 0.56)\times 10^{-5}$ \\
0.600 - 0.800 & 0.697 & 43.61 - 43.73 & 43.67 &  23.0 & $(4.68\pm 0.51)\times 10^{-5}$ \\
0.600 - 0.800 & 0.697 & 43.73 - 43.86 & 43.79 &  14.3 & $(2.08\pm 0.29)\times 10^{-5}$ \\
0.600 - 0.800 & 0.697 & 43.86 - 43.98 & 43.92 &  13.3 & $(1.68\pm 0.25)\times 10^{-5}$ \\
0.600 - 0.800 & 0.697 & 43.98 - 44.13 & 44.06 &   9.5 & $(8.32\pm 1.58)\times 10^{-6}$ \\
0.600 - 0.800 & 0.697 & 44.13 - 44.52 & 44.33 &  12.5 & $(3.20\pm 0.54)\times 10^{-6}$ \\
0.600 - 0.800 & 0.697 & 44.52 - 44.65 & 44.58 &   5.0 & $(1.52\pm 0.53)\times 10^{-6}$ \\
0.600 - 0.800 & 0.697 & 44.65 - 45.08 & 44.87 &  15.5 & $(4.40\pm 0.88)\times 10^{-7}$ \\
0.600 - 0.800 & 0.697 & 45.08 - 45.33 & 45.21 &   0.2 & $(7.45\pm 2.46)\times 10^{-9}$ \\
\\                                                     
0.800 - 1.000 & 0.897 & 41.98 - 42.48 & 42.23 &   3.9 & $(1.49\pm 0.43)\times 10^{-4}$ \\
0.800 - 1.000 & 0.897 & 42.48 - 42.73 & 42.60 &  12.8 & $(3.40\pm 0.51)\times 10^{-4}$ \\
0.800 - 1.000 & 0.897 & 42.73 - 42.85 & 42.79 &  11.5 & $(3.13\pm 0.47)\times 10^{-4}$ \\
0.800 - 1.000 & 0.897 & 42.85 - 42.98 & 42.92 &  15.5 & $(2.27\pm 0.27)\times 10^{-4}$ \\
0.800 - 1.000 & 0.897 & 42.98 - 43.10 & 43.04 &  22.2 & $(2.36\pm 0.26)\times 10^{-4}$ \\
0.800 - 1.000 & 0.897 & 43.10 - 43.23 & 43.17 &  25.3 & $(1.75\pm 0.19)\times 10^{-4}$ \\
0.800 - 1.000 & 0.897 & 43.23 - 43.35 & 43.29 &  25.4 & $(1.26\pm 0.13)\times 10^{-4}$ \\
0.800 - 1.000 & 0.897 & 43.35 - 43.48 & 43.42 &  41.2 & $(1.39\pm 0.11)\times 10^{-4}$ \\
0.800 - 1.000 & 0.897 & 43.48 - 43.60 & 43.54 &  39.3 & $(1.20\pm 0.10)\times 10^{-4}$ \\
0.800 - 1.000 & 0.897 & 43.60 - 43.73 & 43.67 &  39.0 & $(8.68\pm 0.74)\times 10^{-5}$ \\
0.800 - 1.000 & 0.897 & 43.73 - 43.85 & 43.79 &  28.5 & $(5.37\pm 0.54)\times 10^{-5}$ \\
0.800 - 1.000 & 0.897 & 43.85 - 43.98 & 43.92 &  28.2 & $(3.97\pm 0.40)\times 10^{-5}$ \\
0.800 - 1.000 & 0.897 & 43.98 - 44.10 & 44.04 &  16.5 & $(1.93\pm 0.25)\times 10^{-5}$ \\
0.800 - 1.000 & 0.897 & 44.10 - 44.23 & 44.17 &  19.7 & $(1.69\pm 0.22)\times 10^{-5}$ \\
0.800 - 1.000 & 0.897 & 44.23 - 44.48 & 44.35 &  18.3 & $(6.69\pm 0.94)\times 10^{-6}$ \\
0.800 - 1.000 & 0.897 & 44.48 - 44.88 & 44.68 &  10.5 & $(1.50\pm 0.33)\times 10^{-6}$ \\
0.800 - 1.000 & 0.897 & 44.88 - 45.13 & 45.01 &   1.3 & $(6.39\pm 3.90)\times 10^{-8}$ \\
\\                                                     
1.000 - 1.200 & 1.098 & 41.86 - 42.36 & 42.11 &   1.8 & $(1.14\pm 0.42)\times 10^{-4}$ \\
1.000 - 1.200 & 1.098 & 42.36 - 42.85 & 42.60 &  11.1 & $(2.15\pm 0.37)\times 10^{-4}$ \\
1.000 - 1.200 & 1.098 & 42.85 - 42.98 & 42.92 &   6.4 & $(1.76\pm 0.35)\times 10^{-4}$ \\
1.000 - 1.200 & 1.098 & 42.98 - 43.10 & 43.04 &  10.4 & $(1.80\pm 0.29)\times 10^{-4}$ \\
1.000 - 1.200 & 1.098 & 43.10 - 43.23 & 43.17 &  14.2 & $(1.37\pm 0.18)\times 10^{-4}$ \\
1.000 - 1.200 & 1.098 & 43.23 - 43.35 & 43.29 &  14.9 & $(1.12\pm 0.15)\times 10^{-4}$ \\
1.000 - 1.200 & 1.098 & 43.35 - 43.48 & 43.42 &  22.7 & $(1.04\pm 0.11)\times 10^{-4}$ \\
1.000 - 1.200 & 1.098 & 43.48 - 43.60 & 43.54 &  30.4 & $(1.02\pm 0.09)\times 10^{-4}$ \\
1.000 - 1.200 & 1.098 & 43.60 - 43.73 & 43.67 &  31.3 & $(7.51\pm 0.68)\times 10^{-5}$ \\
1.000 - 1.200 & 1.098 & 43.73 - 43.85 & 43.79 &  30.3 & $(6.65\pm 0.63)\times 10^{-5}$ \\
1.000 - 1.200 & 1.098 & 43.85 - 43.98 & 43.92 &  27.6 & $(4.53\pm 0.45)\times 10^{-5}$ \\
1.000 - 1.200 & 1.098 & 43.98 - 44.10 & 44.04 &  24.6 & $(3.66\pm 0.40)\times 10^{-5}$ \\
1.000 - 1.200 & 1.098 & 44.10 - 44.23 & 44.17 &  16.7 & $(1.77\pm 0.23)\times 10^{-5}$ \\
1.000 - 1.200 & 1.098 & 44.23 - 44.35 & 44.29 &  12.6 & $(1.07\pm 0.17)\times 10^{-5}$ \\
1.000 - 1.200 & 1.098 & 44.35 - 44.57 & 44.46 &  11.8 & $(4.37\pm 0.74)\times 10^{-6}$ \\
1.000 - 1.200 & 1.098 & 44.57 - 44.98 & 44.78 &  11.4 & $(1.69\pm 0.34)\times 10^{-6}$ \\
1.000 - 1.200 & 1.098 & 44.98 - 45.34 & 45.16 &   4.3 & $(1.53\pm 0.58)\times 10^{-7}$ \\
1.000 - 1.200 & 1.098 & 45.34 - 45.59 & 45.47 &   4.0 & $(8.73\pm 3.75)\times 10^{-8}$ \\
\\                                                     
1.200 - 1.600 & 1.392 & 42.10 - 42.60 & 42.35 &   7.0 & $(1.82\pm 0.40)\times 10^{-4}$ \\
1.200 - 1.600 & 1.392 & 42.60 - 42.86 & 42.73 &   6.3 & $(1.34\pm 0.31)\times 10^{-4}$ \\
1.200 - 1.600 & 1.392 & 42.86 - 43.04 & 42.95 &   6.7 & $(1.10\pm 0.24)\times 10^{-4}$ \\
1.200 - 1.600 & 1.392 & 43.04 - 43.16 & 43.10 &  10.6 & $(1.53\pm 0.26)\times 10^{-4}$ \\
1.200 - 1.600 & 1.392 & 43.16 - 43.29 & 43.22 &  14.9 & $(1.15\pm 0.15)\times 10^{-4}$ \\
1.200 - 1.600 & 1.392 & 43.29 - 43.41 & 43.35 &  23.1 & $(1.17\pm 0.13)\times 10^{-4}$ \\
1.200 - 1.600 & 1.392 & 43.41 - 43.54 & 43.47 &  26.9 & $(8.54\pm 0.74)\times 10^{-5}$ \\
1.200 - 1.600 & 1.392 & 43.54 - 43.66 & 43.60 &  37.6 & $(8.73\pm 0.71)\times 10^{-5}$ \\
1.200 - 1.600 & 1.392 & 43.66 - 43.79 & 43.72 &  49.2 & $(7.16\pm 0.49)\times 10^{-5}$ \\
1.200 - 1.600 & 1.392 & 43.79 - 43.91 & 43.85 &  50.6 & $(5.98\pm 0.41)\times 10^{-5}$ \\
1.200 - 1.600 & 1.392 & 43.91 - 44.04 & 43.97 &  58.3 & $(5.17\pm 0.34)\times 10^{-5}$ \\
1.200 - 1.600 & 1.392 & 44.04 - 44.16 & 44.10 &  44.0 & $(3.40\pm 0.26)\times 10^{-5}$ \\
1.200 - 1.600 & 1.392 & 44.16 - 44.29 & 44.22 &  39.1 & $(2.31\pm 0.19)\times 10^{-5}$ \\
1.200 - 1.600 & 1.392 & 44.29 - 44.41 & 44.35 &  38.2 & $(2.04\pm 0.18)\times 10^{-5}$ \\
1.200 - 1.600 & 1.392 & 44.41 - 44.54 & 44.47 &  25.8 & $(1.03\pm 0.11)\times 10^{-5}$ \\
1.200 - 1.600 & 1.392 & 44.54 - 44.66 & 44.60 &  23.7 & $(8.19\pm 1.06)\times 10^{-6}$ \\
1.200 - 1.600 & 1.392 & 44.66 - 44.98 & 44.82 &  21.9 & $(2.17\pm 0.26)\times 10^{-6}$ \\
1.200 - 1.600 & 1.392 & 44.98 - 45.10 & 45.04 &   2.0 & $(4.94\pm 2.03)\times 10^{-7}$ \\
1.200 - 1.600 & 1.392 & 45.10 - 45.29 & 45.19 &   4.2 & $(3.90\pm 1.44)\times 10^{-7}$ \\
1.200 - 1.600 & 1.392 & 45.29 - 45.78 & 45.53 &   2.2 & $(1.48\pm 0.77)\times 10^{-8}$ \\
1.200 - 1.600 & 1.392 & 45.78 - 46.03 & 45.90 &   1.0 & $(7.86\pm 7.00)\times 10^{-9}$ \\
\\                                                     
1.600 - 2.000 & 1.793 & 41.91 - 42.41 & 42.16 &   1.3 & $(1.45\pm 0.62)\times 10^{-4}$ \\
1.600 - 2.000 & 1.793 & 42.41 - 42.91 & 42.66 &   7.3 & $(1.38\pm 0.30)\times 10^{-4}$ \\
1.600 - 2.000 & 1.793 & 42.91 - 43.21 & 43.06 &   8.0 & $(1.05\pm 0.22)\times 10^{-4}$ \\
1.600 - 2.000 & 1.793 & 43.21 - 43.38 & 43.29 &   6.5 & $(7.14\pm 1.71)\times 10^{-5}$ \\
1.600 - 2.000 & 1.793 & 43.38 - 43.51 & 43.44 &   9.2 & $(6.95\pm 1.11)\times 10^{-5}$ \\
1.600 - 2.000 & 1.793 & 43.51 - 43.63 & 43.57 &  14.8 & $(7.12\pm 0.85)\times 10^{-5}$ \\
1.600 - 2.000 & 1.793 & 43.63 - 43.76 & 43.69 &  21.8 & $(6.48\pm 0.63)\times 10^{-5}$ \\
1.600 - 2.000 & 1.793 & 43.76 - 43.88 & 43.82 &  22.2 & $(4.85\pm 0.48)\times 10^{-5}$ \\
1.600 - 2.000 & 1.793 & 43.88 - 44.01 & 43.94 &  38.1 & $(5.16\pm 0.37)\times 10^{-5}$ \\
1.600 - 2.000 & 1.793 & 44.01 - 44.13 & 44.07 &  30.5 & $(3.30\pm 0.28)\times 10^{-5}$ \\
1.600 - 2.000 & 1.793 & 44.13 - 44.26 & 44.19 &  38.0 & $(3.07\pm 0.24)\times 10^{-5}$ \\
1.600 - 2.000 & 1.793 & 44.26 - 44.38 & 44.32 &  32.7 & $(2.30\pm 0.20)\times 10^{-5}$ \\
1.600 - 2.000 & 1.793 & 44.38 - 44.51 & 44.44 &  25.3 & $(1.35\pm 0.14)\times 10^{-5}$ \\
1.600 - 2.000 & 1.793 & 44.51 - 44.63 & 44.57 &  18.4 & $(9.15\pm 1.10)\times 10^{-6}$ \\
1.600 - 2.000 & 1.793 & 44.63 - 44.80 & 44.72 &  16.2 & $(4.14\pm 0.54)\times 10^{-6}$ \\
1.600 - 2.000 & 1.793 & 44.80 - 44.93 & 44.86 &   8.3 & $(2.14\pm 0.45)\times 10^{-6}$ \\
1.600 - 2.000 & 1.793 & 44.93 - 45.05 & 44.99 &   3.2 & $(8.10\pm 2.59)\times 10^{-7}$ \\
1.600 - 2.000 & 1.793 & 45.05 - 45.18 & 45.11 &   3.4 & $(7.08\pm 2.55)\times 10^{-7}$ \\
1.600 - 2.000 & 1.793 & 45.18 - 45.61 & 45.40 &   2.1 & $(7.76\pm 2.87)\times 10^{-8}$ \\
1.600 - 2.000 & 1.793 & 45.61 - 45.86 & 45.74 &   0.9 & $(1.14\pm 1.01)\times 10^{-8}$ \\
\\                                                     
2.000 - 2.400 & 2.194 & 42.58 - 43.08 & 42.83 &   4.4 & $(8.19\pm 2.29)\times 10^{-5}$ \\
2.000 - 2.400 & 2.194 & 43.08 - 43.45 & 43.26 &   7.1 & $(6.74\pm 1.55)\times 10^{-5}$ \\
2.000 - 2.400 & 2.194 & 43.45 - 43.63 & 43.54 &   6.9 & $(5.42\pm 1.08)\times 10^{-5}$ \\
2.000 - 2.400 & 2.194 & 43.63 - 43.76 & 43.69 &   9.7 & $(5.26\pm 0.74)\times 10^{-5}$ \\
2.000 - 2.400 & 2.194 & 43.76 - 43.88 & 43.82 &  14.8 & $(5.46\pm 0.66)\times 10^{-5}$ \\
2.000 - 2.400 & 2.194 & 43.88 - 44.01 & 43.94 &  23.2 & $(5.38\pm 0.51)\times 10^{-5}$ \\
2.000 - 2.400 & 2.194 & 44.01 - 44.13 & 44.07 &  30.3 & $(5.15\pm 0.43)\times 10^{-5}$ \\
2.000 - 2.400 & 2.194 & 44.13 - 44.26 & 44.19 &  33.3 & $(3.63\pm 0.30)\times 10^{-5}$ \\
2.000 - 2.400 & 2.194 & 44.26 - 44.38 & 44.32 &  25.3 & $(2.33\pm 0.22)\times 10^{-5}$ \\
2.000 - 2.400 & 2.194 & 44.38 - 44.51 & 44.44 &  27.4 & $(1.88\pm 0.18)\times 10^{-5}$ \\
2.000 - 2.400 & 2.194 & 44.51 - 44.63 & 44.57 &  19.9 & $(1.19\pm 0.14)\times 10^{-5}$ \\
2.000 - 2.400 & 2.194 & 44.63 - 44.80 & 44.72 &  17.0 & $(5.97\pm 0.78)\times 10^{-6}$ \\
2.000 - 2.400 & 2.194 & 44.80 - 44.96 & 44.88 &   9.0 & $(2.44\pm 0.44)\times 10^{-6}$ \\
2.000 - 2.400 & 2.194 & 44.96 - 45.15 & 45.06 &   3.7 & $(6.00\pm 1.44)\times 10^{-7}$ \\
2.000 - 2.400 & 2.194 & 45.15 - 45.39 & 45.27 &   2.8 & $(2.94\pm 1.09)\times 10^{-7}$ \\
2.000 - 2.400 & 2.194 & 45.39 - 45.77 & 45.58 &   0.3 & $(1.11\pm 0.51)\times 10^{-8}$ \\
2.000 - 2.400 & 2.194 & 45.77 - 46.02 & 45.90 &   0.9 & $(1.22\pm 1.10)\times 10^{-8}$ \\
\\                                                     
2.400 - 3.000 & 2.688 & 42.71 - 43.21 & 42.96 &   3.6 & $(5.25\pm 1.68)\times 10^{-5}$ \\
2.400 - 3.000 & 2.688 & 43.21 - 43.59 & 43.40 &   9.0 & $(6.24\pm 1.12)\times 10^{-5}$ \\
2.400 - 3.000 & 2.688 & 43.59 - 43.72 & 43.65 &   8.5 & $(8.04\pm 1.61)\times 10^{-5}$ \\
2.400 - 3.000 & 2.688 & 43.72 - 43.84 & 43.78 &   7.4 & $(4.45\pm 0.89)\times 10^{-5}$ \\
2.400 - 3.000 & 2.688 & 43.84 - 43.97 & 43.90 &  13.8 & $(4.36\pm 0.57)\times 10^{-5}$ \\
2.400 - 3.000 & 2.688 & 43.97 - 44.09 & 44.03 &  14.0 & $(3.09\pm 0.37)\times 10^{-5}$ \\
2.400 - 3.000 & 2.688 & 44.09 - 44.22 & 44.15 &  22.2 & $(3.07\pm 0.30)\times 10^{-5}$ \\
2.400 - 3.000 & 2.688 & 44.22 - 44.34 & 44.28 &  19.3 & $(1.98\pm 0.22)\times 10^{-5}$ \\
2.400 - 3.000 & 2.688 & 44.34 - 44.47 & 44.40 &  23.9 & $(1.61\pm 0.16)\times 10^{-5}$ \\
2.400 - 3.000 & 2.688 & 44.47 - 44.59 & 44.53 &  21.8 & $(1.25\pm 0.14)\times 10^{-5}$ \\
2.400 - 3.000 & 2.688 & 44.59 - 44.74 & 44.67 &  18.6 & $(6.80\pm 0.75)\times 10^{-6}$ \\
2.400 - 3.000 & 2.688 & 44.74 - 44.87 & 44.81 &  12.3 & $(4.16\pm 0.62)\times 10^{-6}$ \\
2.400 - 3.000 & 2.688 & 44.87 - 45.33 & 45.10 &  14.2 & $(7.08\pm 1.06)\times 10^{-7}$ \\
2.400 - 3.000 & 2.688 & 45.33 - 45.58 & 45.46 &   0.6 & $(3.99\pm 1.36)\times 10^{-8}$ \\
\\                                                     
3.000 - 5.800 & 4.215 & 42.97 - 43.47 & 43.22 &   2.9 & $(7.94\pm 2.94)\times 10^{-6}$ \\
3.000 - 5.800 & 4.215 & 43.47 - 43.75 & 43.61 &   3.7 & $(7.03\pm 1.76)\times 10^{-6}$ \\
3.000 - 5.800 & 4.215 & 43.75 - 43.97 & 43.86 &   8.5 & $(9.67\pm 1.74)\times 10^{-6}$ \\
3.000 - 5.800 & 4.215 & 43.97 - 44.10 & 44.03 &   5.9 & $(5.56\pm 1.11)\times 10^{-6}$ \\
3.000 - 5.800 & 4.215 & 44.10 - 44.28 & 44.19 &  13.2 & $(4.85\pm 0.68)\times 10^{-6}$ \\
3.000 - 5.800 & 4.215 & 44.28 - 44.41 & 44.35 &   5.5 & $(1.70\pm 0.31)\times 10^{-6}$ \\
3.000 - 5.800 & 4.215 & 44.41 - 44.53 & 44.47 &   8.3 & $(1.87\pm 0.30)\times 10^{-6}$ \\
3.000 - 5.800 & 4.215 & 44.53 - 44.75 & 44.64 &  19.5 & $(1.55\pm 0.19)\times 10^{-6}$ \\
3.000 - 5.800 & 4.215 & 44.75 - 45.04 & 44.90 &  14.4 & $(5.38\pm 0.70)\times 10^{-7}$ \\
3.000 - 5.800 & 4.215 & 45.04 - 45.29 & 45.17 &   4.9 & $(1.25\pm 0.34)\times 10^{-7}$ \\
\enddata
\tablenotetext{a}{Units -- $\frac{d \Phi}{d\log \, L_{\rm X}}$: [$h_{70}^3\;{\rm Mpc^{-3}dex^{-1}}$],\,\,
$L_{\rm X}$: $[\;h_{70}^{-2}{\rm erg\;s^{-1}}]$.}
\tablenotetext{b}{Errors show the 68\% confidence range for one parameter ($\Delta \mathcal{L}<1$).} 
\end{deluxetable}

\begin{figure}
\begin{center}
\resizebox{!}{!}{
  \includegraphics[width=\hsize]{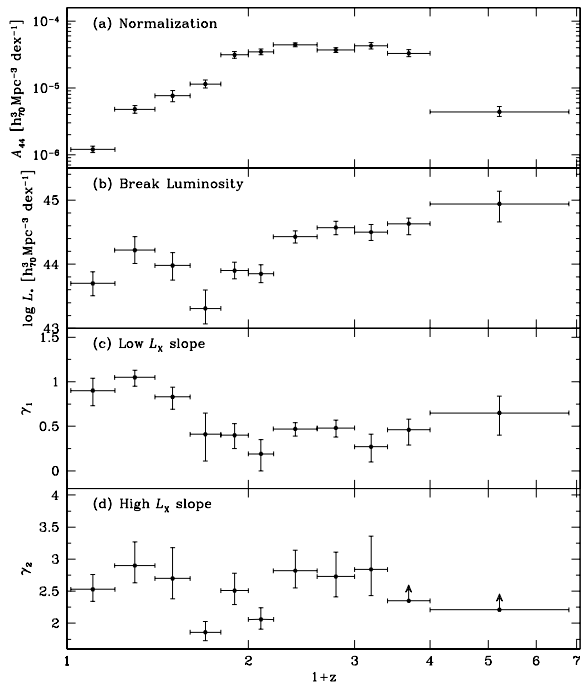}
}
\end{center}

\caption{The parameters of the smoothed two power-law fits to individual 
redshift shells are plotted as a function of redshift. See labels. The error bars
are 1$\sigma$ corresponding to $\Delta L=1$. The upward arrow of in panel (d) 
corresponds to a 90\% lower limit of $\gamma_2$.}
\label{fig:param_zev}
\end{figure}

Figure \ref{fig:param_zev} shows the variation of the smoothed two-power law parameters
as a function of redshift. The normalization is the XLF at $\log L_{\rm X}=44$ and we see the familiar 
feature of growth (with increasing redshift) between $z=0$ and $z\sim 1$, followed by a plateau, then a decline at $z>3$. 
A notable feature is a sudden drop of the low luminosity slope $\gamma_1$ from $z< 0.6$ ($\gamma_1\sim 1$), to 
$z\geq 0.6$ ($\gamma \sim 0.5$). No significant change of slope is observed in each side of this redshift. 
An F-test comparison between the best-fit models for the case where $\gamma_1$ is constant over redshift versus the case 
where $\gamma_1$ changes at $z=0.6$ shows that the probability that the former is accepted is only $7\times 10^{-5}$.   
 The high luminosity slope is consistent with being constant at $\gamma_2\sim 2.7$, except for 
two data points at $z\sim 0.7$ and $z\sim 1.1$, where $\gamma_2$ is significantly smaller,
although it may be caused by an observational bias as discussed in Sect. \ref{sec:disc}.  
 
\subsection{Evolution of the Space Density in Luminosity Classes}
\label{sec:evol}
 
 In this section, we investigate the evolution of the AGN space density in different luminosity 
classes as a function of redshift. We have made the fit to an analytical expression at the center 
of each of the four luminosity classes: 42.0--43.0, 43.0--44.0, 44.0--45.0, and 45.0--46.0 in 
$\log L_{\:rm x}$. The XLF at the central luminosity of the bin $\log L_{\rm c}$, which is defined 
as the mean of the minimum and maximum bounds of $\log L_{\rm x}$ bin can be expressed as:
\begin{equation}
\frac{{\rm d}\;\Phi\,(L_{\rm c},z)}{{\rm d\;log}\;L_{\rm X}}
  = A_{L_{\rm c}}^{z=0}\cdot e_{\rm d}(z, L_{\rm c}),
\label{eq:ldde0l}
\end{equation}
where the normalization $A_{L_{\rm c}}^{z=0}$ is the XLF value at the central luminosity of 
the class at $z=0$.

 For each luminosity class, we have made a maximum-likelihood fit to the 
the redshift-dependent evolution parameters in the evolution factor expressed as
a three-segment power-law model Eq. \ref{eq:3po_ev}, evaluated at $L_{\rm X}=L_{\rm c}$.

  During the fitting process, the luminosity dependence of the XLF within the shell is fixed to the 
best-fit LDDE case. In each class, there are six fitting parameters:  $A_{L_{\rm c}}^{z=0}$, $p_1$, $z_{\rm b1}$, 
$p_2$, $z_{\rm b2}$ and $p_3$. As in the case of redshift shell divided samples, maximum-likelihood fits have been made 
for each luminosity class using Eq. \ref{eq:ml1}, where the normalization is also a fitting parameter.
Thus the reported normalization errors contain the effects of correlations with other parameters.
The resulting best-fit parameters and errors are reported in Table \ref{tab:lldv_par}.


 One limitation of dividing the sample into luminosity classes is that we do not know the 
intrinsic luminosity $L_{\rm X}$ of each object. We only know the observed flux and redshift.
For the fit of the luminosity class $(\log L_{\rm X,min},\log L_{\rm X,max})$, we select  
objects that fall into the $\log L_{\rm X,min}\leq \log L_{\rm X,obs}\leq \log L_{\rm X,max}$ 
range. Thus the fitting in each luminosity class uses some objects that are outside of the 
$\log L_{\rm X,min}\leq \log L_{\rm X,obs}\leq \log L_{\rm X,max}$ range, while others that fall into
this range are not used. The fitting process using Eq. \ref{eq:n_func_abs} properly takes care of 
the expected number of AGNs in the observed luminosity range calculated based on the intrinsic luminosity XLF model.

\begin{deluxetable}{cccccccc}
\tabletypesize{\footnotesize}
\tablecaption{Best--fit evolution parameters for each luminosity class\label{tab:lldv_par}}
\tablewidth{0pt}
\tablehead{
\colhead{$\log L_{\rm X}$-range\tablenotemark{a}} & \colhead{$\log L_{\rm c}$\tablenotemark{a}} & \colhead{$A_{L{\rm c}}^{z=0 }$\tablenotemark{a}}
 & \colhead{$p_1$\tablenotemark{a}} & \colhead{$z_{\rm b1}$\tablenotemark{a}} &  \colhead{$p_2$\tablenotemark{a}} 
 & \colhead{$z_{\rm b2}$\tablenotemark{a}} & \colhead{$p_3$\tablenotemark{a}}}
\startdata
  42.0 -  43.0 &  42.5 & $ (8.3\pm 1.0)\times 10^{-5}$ & $  3.7\pm 0.3$ & $ 0.65^{+0.04}_{-0.03}$ & $ -3.4^{+0.5}_{-0.6}$ & ... & ...\\
  43.0 -  44.0 &  43.5 & $ (6.0\pm 0.4)\times 10^{-6}$ & $  4.7\pm 0.2$ & $ 0.85^{+0.03}_{-0.02}$ & $ -1.0^{+0.2}_{-0.2}$ &  $ 2.69^{+0.10}_{-0.18}$ & $ -5.9^{+0.0}_{-1.7}$\\
  44.0 -  45.0 &  44.5 & $ (6.8\pm 0.9)\times 10^{-8}$ & $  5.8^{+0.2}_{-0.3}$ & $ 1.41^{+0.09}_{-0.11}$ & $  0.4^{+0.3}_{-0.5}$ &  $ 2.63^{+0.16}_{-0.11}$ & $ -5.4^{+0.9}_{-1.1}$\\
  45.0 -  46.0 &  45.5 & $ (8^{+6}_{-4})\times 10^{-11}$ & $ 7.3\pm 0.8$ & $ 1.40$* & $  2.2\pm 0.9$ &  $ 2.6$* & $ -4.7^{+1.2}_{-1.3}$\\
\enddata
\tablenotetext{a}{The same units and error definitions as in Table \ref{tab:zshell2po} apply.}
\tablenotetext{b}{Fixed parameters are indicated by an asterisk ('*').}
\end{deluxetable}
 
\begin{figure*}
\begin{center}
\resizebox{\hsize}{!}{\includegraphics[width=\hsize]{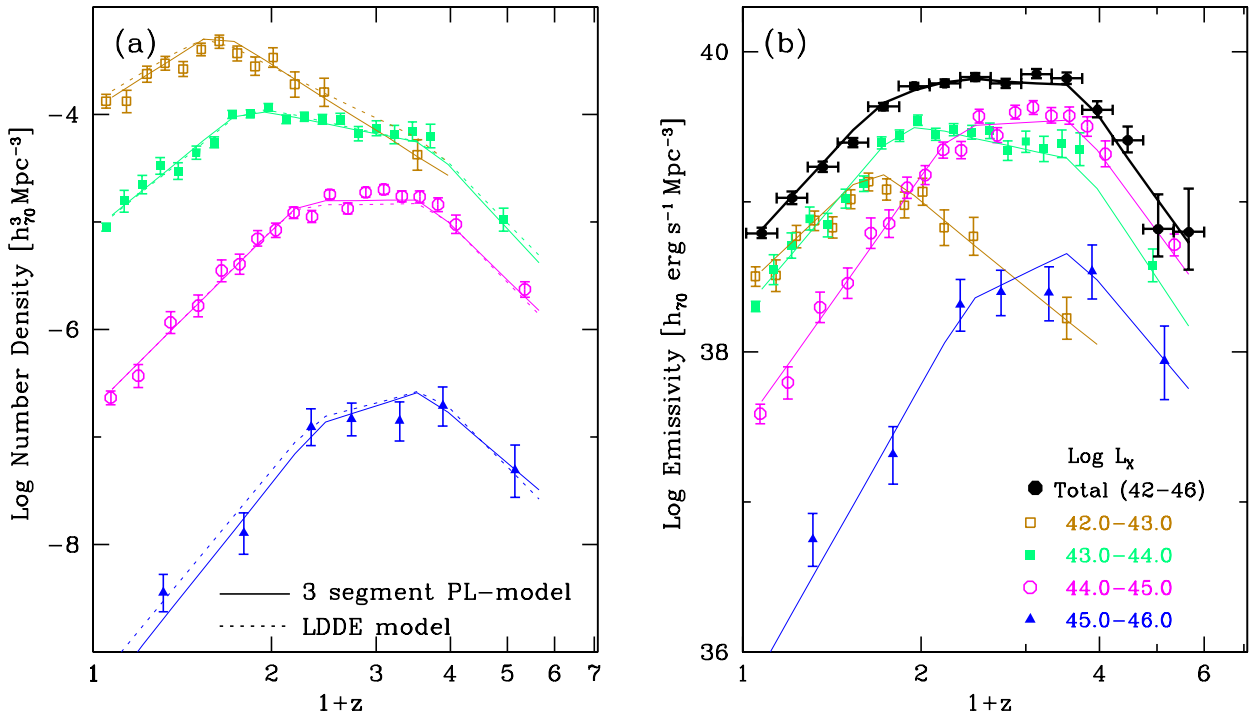}}
\end{center}
\caption{(a) The space density of AGNs as a function of redshift in 
different luminosity classes as labeled in panel (b).  Densities from the luminosity-class
divided fits as well as those from the LDDE models are 
overplotted with solid lines and dotted lines respectively.
(b) The 2-10 keV X--ray emissivities are plotted instead of number densities. Also
the total emissivity curves are added. Only models from the luminosity-divided fits 
are plotted in this panel. (Color versions of these figures are available in the electronic
version.)} 
\label{fig:ndens}
\end{figure*}

 Figure \ref{fig:ndens} shows the evolution of the AGN number density and emissivity in 
each class for the full treatment.  The binned number densities and emissivities, calculated with 
the $N_{\rm obs}/N_{\rm mdl}$  method using the luminosity-class divided models, are shown with 
68\% Poisson errors.
The total emissivities of the $42\leq \log L_{\rm X} [h_{\rm 70}\,erg\,s^{-1}]<46$ range are also
plotted as a function of redshift. The model number densities and emissivities are also
shown for the best-fit LDDE model in Table \ref{tab:best} as well as class-divided fits. The
evolution of total emissivity in $42<\log L_{\rm X}<46$ is also plotted. Table \ref{tab:exlf_llclass}
tabulates the binned $N_{\rm obs}/N_{\rm mdl}$-estimated values evaluated at the central luminosities $L_{\rm c}$ 
(instead of number densities integrated over the luminosity class, which are plotted in Fig. \ref{fig:ndens}), 
as well as emissivities with best fit models.
  
The results in Table \ref{tab:lldv_par} show that the first break redshift $z_{\rm b1}$ increases with luminosity 
with $z_{\rm b1}\sim 1.1$ at $\log L_{\rm X}\approx 44 {\rm [h_{\rm 70}\,erg\,s^{-1}]}$.
The second break { redshift does not vary significantly} and is consistent with $z_{\rm b2}\sim 2.7$ at 
$\log L_{\rm X}>43\;[h_{\rm 70}{\rm \,erg\,s^{-1}}]$. The slope after the second break is consistent with  $p_3\approx -5.6$. 
These have been fed back as fixed parameters for a refinement of the global expression in Sect. \ref{sec:xlf_global}.   
{ For the $42\le \log L_{\rm X}<43$ class, the values of $z_{\rm b2}$ and $p_3$ could not be determined, because the 
sample does not extend to the second break $z_{\rm b2}\sim 2.7$. For the  $45\le \log L_{\rm X}<46$ class, 
the low number of objects has limited our analysis. While we obtain best fit values of $z_{\rm b1}\approx 1.5$ and  
$z_{\rm b2}\approx 3$, their 1$\sigma$ errors are essentially unconstrained.
Thus we fixed the $z_{\rm b1}=1.2$ and $z_{\rm b2}=2.6$ based on the results of the fit to the $44\le \log L_{\rm X}<45$ 
class.  We have also attempted a two-segment law fit (one break redshift) for this class, but the three segment form 
with the two break redshifts fixed at these values gives a better fit.}    

While the value of the parameter $p_1$ increases with luminosity, the measurements are over different redshift 
ranges for different luminosity classes and should not be compared with one another. The $p_2$ parameters are 
also measured over somewhat different redshift ranges. However, it is clear from Fig. \ref{fig:ndens} that
the slope between $z_{\rm b1}$ and $z_{\rm b2}$ goes from negative to positive (as a function of $\log (1+z)$)
with luminosity.   The full binned  XLF divided in luminosity classes is shown in Table~\ref{tab:exlf_llclass}.
 
\begin{deluxetable}{cccccc}
\tabletypesize{\footnotesize}
\tablecaption{Full binned XLF divided in luminosity classes\label{tab:exlf_llclass}}
\tablewidth{0pt}
\tablehead{
\colhead{$z$-range} &\colhead{$z_{\rm c}$} &\colhead{$\log L_{\rm X}$-range\tablenotemark{a}} & \colhead{$\log L_{\rm X,c}$\tablenotemark{a}} & \colhead{$N_{\rm obj}$} & \colhead{$\frac{d\Phi}{d\,L_{\rm X}}$\tablenotemark{a,b}} 
}
\startdata 
0.015 - 0.089 & 0.051 & 42.00 - 43.00 & 42.50 &  51.6 &$(1.01\pm 0.13)\times 10^{-4}$ \\
0.089 - 0.190 & 0.138 & 42.00 - 43.00 & 42.50 &  21.2 &$(1.02\pm 0.13)\times 10^{-4}$ \\
0.190 - 0.276 & 0.232 & 42.00 - 43.00 & 42.50 &  38.6 &$(1.86\pm 0.18)\times 10^{-4}$ \\
0.276 - 0.369 & 0.322 & 42.00 - 43.00 & 42.50 &  51.4 &$(2.38\pm 0.20)\times 10^{-4}$ \\
0.369 - 0.468 & 0.418 & 42.00 - 43.00 & 42.50 &  42.9 &$(2.13\pm 0.19)\times 10^{-4}$ \\
0.468 - 0.575 & 0.521 & 42.00 - 43.00 & 42.50 &  57.3 &$(3.29\pm 0.25)\times 10^{-4}$ \\
0.575 - 0.689 & 0.631 & 42.00 - 43.00 & 42.50 &  61.3 &$(4.67\pm 0.34)\times 10^{-4}$ \\
0.689 - 0.811 & 0.749 & 42.00 - 43.00 & 42.50 &  45.6 &$(3.60\pm 0.29)\times 10^{-4}$ \\
0.811 - 0.942 & 0.875 & 42.00 - 43.00 & 42.50 &  29.2 &$(2.70\pm 0.26)\times 10^{-4}$ \\
0.942 - 1.083 & 1.011 & 42.00 - 43.00 & 42.50 &  27.5 &$(3.30\pm 0.33)\times 10^{-4}$ \\
1.083 - 1.299 & 1.188 & 42.00 - 43.00 & 42.50 &  16.9 &$(1.86\pm 0.26)\times 10^{-4}$ \\
1.299 - 1.615 & 1.452 & 42.00 - 43.00 & 42.50 &  14.7 &$(1.59\pm 0.24)\times 10^{-4}$ \\
1.615 - 3.744 & 2.522 & 42.00 - 43.00 & 42.50 &  12.0 &$(4.18\pm 0.67)\times 10^{-5}$ \\
\\                                                    
0.015 - 0.089 & 0.051 & 43.00 - 44.00 & 43.50 & 165.3 &$(6.56\pm 0.51)\times 10^{-6}$ \\
0.089 - 0.168 & 0.128 & 43.00 - 44.00 & 43.50 &  22.7 &$(1.16\pm 0.15)\times 10^{-5}$ \\
0.168 - 0.252 & 0.209 & 43.00 - 44.00 & 43.50 &  30.3 &$(1.69\pm 0.19)\times 10^{-5}$ \\
0.252 - 0.343 & 0.297 & 43.00 - 44.00 & 43.50 &  39.5 &$(2.55\pm 0.24)\times 10^{-5}$ \\
0.343 - 0.440 & 0.391 & 43.00 - 44.00 & 43.50 &  34.0 &$(2.33\pm 0.22)\times 10^{-5}$ \\
0.440 - 0.545 & 0.492 & 43.00 - 44.00 & 43.50 &  56.4 &$(3.50\pm 0.26)\times 10^{-5}$ \\
0.545 - 0.657 & 0.600 & 43.00 - 44.00 & 43.50 &  75.7 &$(4.42\pm 0.28)\times 10^{-5}$ \\
0.657 - 0.777 & 0.716 & 43.00 - 44.00 & 43.50 & 140.4 &$(8.33\pm 0.37)\times 10^{-5}$ \\
0.777 - 0.906 & 0.840 & 43.00 - 44.00 & 43.50 & 147.0 &$(1.03\pm 0.05)\times 10^{-4}$ \\
0.906 - 1.044 & 0.974 & 43.00 - 44.00 & 43.50 & 171.3 &$(1.16\pm 0.05)\times 10^{-4}$ \\
1.044 - 1.192 & 1.117 & 43.00 - 44.00 & 43.50 & 126.2 &$(9.00\pm 0.40)\times 10^{-5}$ \\
1.192 - 1.351 & 1.270 & 43.00 - 44.00 & 43.50 & 119.2 &$(9.54\pm 0.44)\times 10^{-5}$ \\
1.351 - 1.522 & 1.435 & 43.00 - 44.00 & 43.50 &  99.1 &$(9.13\pm 0.43)\times 10^{-5}$ \\
1.522 - 1.704 & 1.611 & 43.00 - 44.00 & 43.50 &  83.4 &$(9.25\pm 0.50)\times 10^{-5}$ \\
1.704 - 1.901 & 1.801 & 43.00 - 44.00 & 43.50 &  49.9 &$(6.77\pm 0.43)\times 10^{-5}$ \\
1.901 - 2.111 & 2.004 & 43.00 - 44.00 & 43.50 &  44.7 &$(7.66\pm 0.55)\times 10^{-5}$ \\
2.111 - 2.336 & 2.222 & 43.00 - 44.00 & 43.50 &  32.2 &$(6.79\pm 0.59)\times 10^{-5}$ \\
2.336 - 2.578 & 2.455 & 43.00 - 44.00 & 43.50 &  27.7 &$(7.25\pm 0.70)\times 10^{-5}$ \\
2.578 - 2.838 & 2.706 & 43.00 - 44.00 & 43.50 &  19.5 &$(6.56\pm 0.79)\times 10^{-5}$ \\
2.838 - 5.309 & 3.921 & 43.00 - 44.00 & 43.50 &  20.1 &$(1.10\pm 0.13)\times 10^{-5}$ \\
\\                                                    
0.033 - 0.108 & 0.070 & 44.00 - 45.00 & 44.50 &  52.1 &$(8.71\pm 1.22)\times 10^{-8}$ \\
0.108 - 0.282 & 0.192 & 44.00 - 45.00 & 44.50 &  20.3 &$(1.45\pm 0.29)\times 10^{-7}$ \\
0.282 - 0.422 & 0.350 & 44.00 - 45.00 & 44.50 &  21.9 &$(4.74\pm 0.76)\times 10^{-7}$ \\
0.422 - 0.590 & 0.504 & 44.00 - 45.00 & 44.50 &  22.2 &$(7.05\pm 1.06)\times 10^{-7}$ \\
0.590 - 0.705 & 0.646 & 44.00 - 45.00 & 44.50 &  22.4 &$(1.55\pm 0.23)\times 10^{-6}$ \\
0.705 - 0.829 & 0.766 & 44.00 - 45.00 & 44.50 &  25.1 &$(1.82\pm 0.24)\times 10^{-6}$ \\
0.829 - 0.962 & 0.894 & 44.00 - 45.00 & 44.50 &  43.4 &$(3.20\pm 0.29)\times 10^{-6}$ \\
0.962 - 1.104 & 1.032 & 44.00 - 45.00 & 44.50 &  50.5 &$(3.96\pm 0.32)\times 10^{-6}$ \\
1.104 - 1.256 & 1.179 & 44.00 - 45.00 & 44.50 &  71.3 &$(5.95\pm 0.40)\times 10^{-6}$ \\
1.256 - 1.420 & 1.337 & 44.00 - 45.00 & 44.50 &  67.9 &$(7.03\pm 0.46)\times 10^{-6}$ \\
1.420 - 1.595 & 1.506 & 44.00 - 45.00 & 44.50 & 107.7 &$(1.33\pm 0.07)\times 10^{-5}$ \\
1.595 - 1.784 & 1.688 & 44.00 - 45.00 & 44.50 &  76.6 &$(1.00\pm 0.06)\times 10^{-5}$ \\
1.784 - 1.985 & 1.883 & 44.00 - 45.00 & 44.50 & 101.5 &$(1.45\pm 0.07)\times 10^{-5}$ \\
1.985 - 2.202 & 2.092 & 44.00 - 45.00 & 44.50 &  99.7 &$(1.57\pm 0.08)\times 10^{-5}$ \\
2.202 - 2.434 & 2.316 & 44.00 - 45.00 & 44.50 &  78.2 &$(1.40\pm 0.08)\times 10^{-5}$ \\
2.434 - 2.683 & 2.556 & 44.00 - 45.00 & 44.50 &  68.3 &$(1.41\pm 0.08)\times 10^{-5}$ \\
2.683 - 2.950 & 2.814 & 44.00 - 45.00 & 44.50 &  51.7 &$(1.20\pm 0.08)\times 10^{-5}$ \\
2.950 - 3.236 & 3.091 & 44.00 - 45.00 & 44.50 &  29.2 &$(7.85\pm 0.73)\times 10^{-6}$ \\
3.236 - 5.749 & 4.347 & 44.00 - 45.00 & 44.50 &  40.1 &$(1.96\pm 0.16)\times 10^{-6}$ \\
5.749 - 7.666 & 6.648 & 44.00 - 45.00 & 44.50 &   0.0 &$< 1.1\times 10^{-6}$ \\
\\                                                    
0.173 - 0.470 & 0.313 & 45.00 - 46.00 & 45.50 &   8.1 &$(1.03\pm 0.36)\times 10^{-9}$ \\
0.470 - 1.190 & 0.794 & 45.00 - 46.00 & 45.50 &   6.9 &$(4.18\pm 1.21)\times 10^{-9}$ \\
1.190 - 1.480 & 1.330 & 45.00 - 46.00 & 45.50 &   8.6 &$(4.45\pm 1.20)\times 10^{-8}$ \\
1.480 - 2.000 & 1.728 & 45.00 - 46.00 & 45.50 &  10.5 &$(5.67\pm 1.13)\times 10^{-8}$ \\
2.000 - 2.600 & 2.286 & 45.00 - 46.00 & 45.50 &   7.9 &$(5.93\pm 1.19)\times 10^{-8}$ \\
2.600 - 3.200 & 2.888 & 45.00 - 46.00 & 45.50 &   7.7 &$(8.42\pm 1.77)\times 10^{-8}$ \\
3.200 - 5.310 & 4.148 & 45.00 - 46.00 & 45.50 &   4.6 &$(2.13\pm 0.55)\times 10^{-8}$ \\
\enddata
\tablenotetext{a}{Units -- $\frac{d \Phi}{d\log \, L_{\rm X}}$: [$h_{70}^3\;{\rm Mpc^{-3}dex^{-1}}$],\,\,
$L_{\rm X}$: $[\;h_{70}^{-2}{\rm erg\;s^{-1}}]$.}
\tablenotetext{b}{Errors show the 68\% confidence range for one parameter ($\Delta \mathcal{L}<1$). The symbol '$<$' shows a 90\% upper limit.}
\end{deluxetable}

 Figure~\ref{fig:ndens}(a) clearly shows a shift of the number density peak with luminosity up to 
$\log L_{\rm X}\approx 44.5$ in the sense that more luminous AGNs (QSOs) peak earlier in the history of the universe, 
while the low luminosity ones arise later. Also, there is a clear decline of the derived space densities at all
luminosity classes. A notable feature is a clear plateau structure at intermediate redshifts at intermediate luminosities, which 
we have been able to trace in detail thanks to the addition of the COSMOS dataset. This has been reflected in the total 
emissivity curve, which stays almost constant between $1<z<2.7$.  The slope between the first break $z_{\rm b1}$ 
and the second break $z_{\rm b2}$ changes from negative to positive. 

\section{Discussion}
\label{sec:disc}

\subsection{Sensitivity to pz-PDZs and Spectral Models}
\label{sec:pdf_spec}

\begin{figure*}
\begin{center}
\resizebox{!}{!}{
  \includegraphics[width=0.8\hsize]{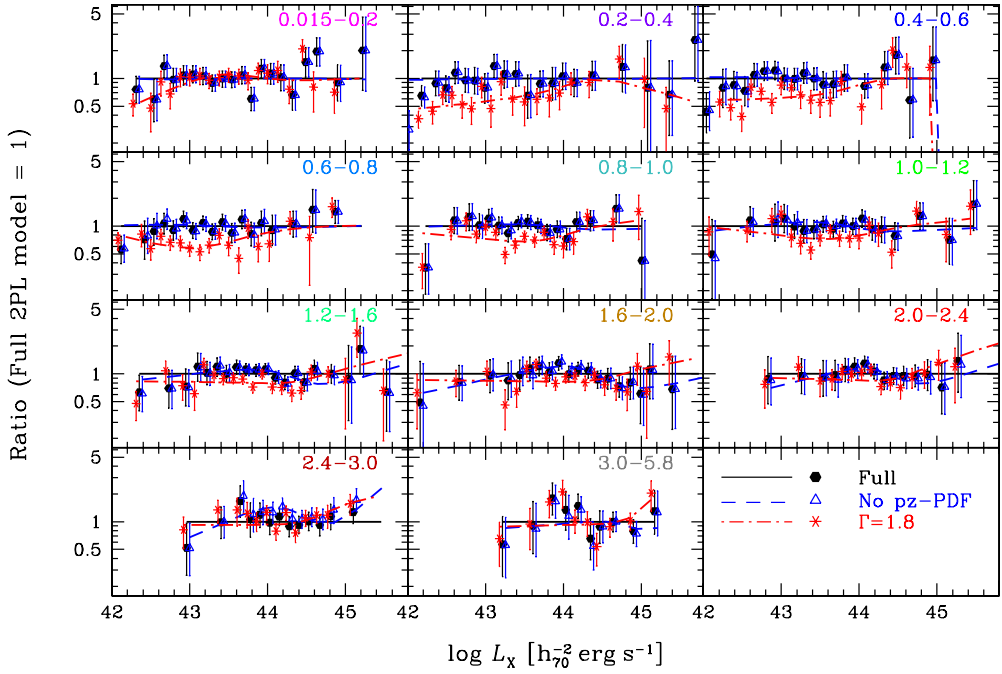}
}
\end{center}
\caption{The comparison of best fit models in each redshift shell and the $N_{\rm obs}/N_{\rm mdl}$ estimates. The best fit two power-law 
models of the case of the no pz-PDFs and $\Gamma=1.8$ are divided by those of the full case. 
The  $N_{\rm obs}/N_{\rm mdl}$ XLFs are also divided by the full model. See labels in the lowest rightmost panel for the meanings of line styles, 
symbols, and colors (electronic version). The data points of the no PDF and $\Gamma=1.8$ cases are shifted horizontally by $+0.04$ 
and $-0.04$ respectively  in $\log L_{\rm X}$ for display purposes. (Color version of this figure is available in the electronic
version.)}
\label{fig:compare_model_z}
\end{figure*}

\begin{figure}
\begin{center}
\resizebox{!}{!}{
  \includegraphics[width=\hsize]{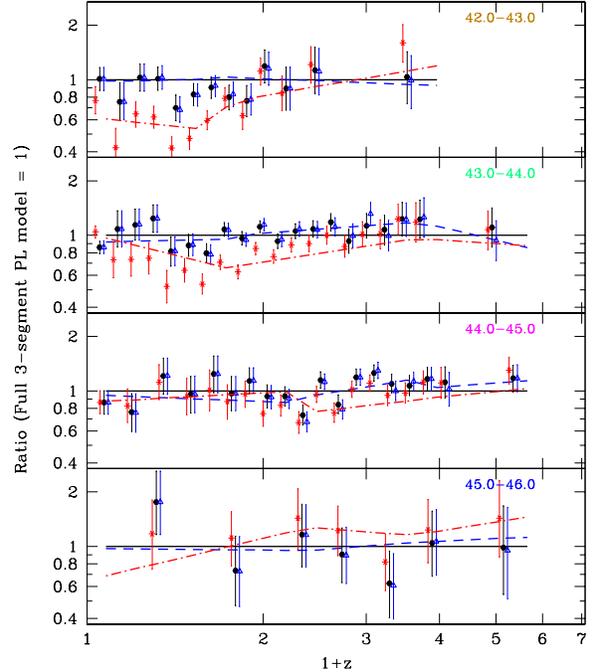}
}
\end{center}
\caption{The comparison of best fit models and the $N_{\rm obs}/N_{\rm mdl}$ estimates in each luminosity class. The best fit three
segment power-law models of No pz-PDFs and $\Gamma=1.8$ cases are divided by those of the full case. 
The  $N_{\rm obs}/N_{\rm mdl}$ XLFs are also divided by the full model. 
The data points of the no pz-PDF and $\Gamma=1.8$ cases are shifted horizontally by $+0.007$ and $-0.007$ respectively in $\log (1+z)$ 
for display purposes. See Fig. \ref{fig:compare_model_z} for the meanings of line styles, symbols, and colors.
(Color version of this figure is available in the electronic
version.)}
\label{fig:compare_model_ll}
\end{figure}

  For comparison, we show the parameters for the case where we only take the best-fit single photometric redshift (PZ) 
per object, instead of full pz-PDFs, under ``No PDZ'' for the global expression (Table \ref{tab:best}). The fitted 
parameters are not significantly different between the two cases except $p2_{44}$, which is the slope between the
first break and the second break redshifts at $\log L_{\rm X}=44$. { The difference in the normalizations are
more than the errors given, but these errors are underestimates, since, for the global expression, we use Eq. \ref{eq:ml2} 
and  correlations with other parameters have not been taken into account, as explained in Sect. \ref{sec:ml12} and 
\ref{sec:ldde_res}}. To observe the effects more closely, we plot the ratio of the XLF estimates based on the full PDF and no PDF samples
for each redshift shell in Fig. \ref{fig:compare_model_z}. The same are plotted for the number density evolution in 
each luminosity class in Fig. \ref{fig:compare_model_ll}. { Upon the examining the difference between the 
best-fit models in these figures, the effects of the pz-PDF seem to be  as large as $\sim 30\%$ at some locations (blue dashed lines). 
However, comparing the data points from the $N_{\rm obs}/N_{\rm mdl}$ (black filled hexagonal versus blue open triangle data points 
in these figures), we see that there is no significant difference and the apparent discrepancy of the models are at locations that
are not well-constrained by the data.  Therefore the apparent deviation is not caused by the pz-PDFs, but by the different convergence of  
of the model parameters at the level of a fraction of statistical errors. This is an optimistic case, because in the COSMOS survey, 
where the majority of the photometric redshifts with the pz-PDZ came from, has exceptionally high quality photometric data, achieving  
an accuracy of $\Delta z/(1+z)\approx 0.015$  with $\sim 5\%$ outliers \citep{salvato11}. At the time of our analysis, full pz-PDFs were not 
available for the CLANS, CLASXS, CDF-N and CDF-S data in our disposal, in which the spectroscopic redshift completeness ranges from 60-80\%. 
These fractions are similar to those of the LH and COSMOS dataset.  For these, we take best-fit photometric redshift values, or for 
CDF-S sources, if the information about the second peak is available, \citep{luo10,xue11}, we give a weight of 0.5 in each peak for the full analysis. 
The photometric redshifts of the CDF-S have a normalized median absolute deviation (NAMD) of $\sigma_{\rm NMAD}=0.059$ with outliers 
of 8.6\% \citep{luo10}, indicating a similar quality to those of COSMOS. Those of the CLANS, CLASXS and CDF-N surveys use fewer photometric 
bands and the uncertainties are as large as $\sigma_{\rm z}/(1+z)\sim 0.16$, but with $\sim 6$\% outliers \citep{trouille08}. The uncertainties
of these PZ have not been taken into account in our full analysis. For a further test,
we create fake PDZ's for objects with photometric redshifts (except those with a second peak) in these surveys with six delta functions.
We assume a Gaussian profile centered at the cataloged PZ value $z_{\rm p}$ for the main pz-PDF peak with an appropriate  
width corresponding to the $\sigma_{\rm NAMD}$ or $\sigma$ values for the each survey. In our fake PDZ, the main peak is represented by three
delta functions, one at the center (representing the central 1/2 of the PDF main peak), the others at the 12.5\% and 87.5\% percentiles of the 
profile (representing the upper and lower 1/4 of the PDF main peak). The central one has a 
weight that is half of $0.5(1 - \mathrm{outlier fraction})$ and the other two each has a weight that is half of the central one. 
Also added are three delta function to represent outliers.  Each of the three  has a weight of 1/3 of the outlier fraction and the redshift 
is taken randomly from spectroscopic redshifts of X-ray sources that have fluxes within a factor of three of that the object. 
We keep using the real PDZs for the COSMOS and LH samples. We make global fit the sample and all the fitted parameters are within a fraction of 
1$\sigma$ errors from the original full analysis.  Thus we conclude that the uncertainties of the PZ's have minimal net effect to our
XLF analysis.}    
   
 In the literature, it has been often assumed for the ``hard''  ($2 \le E{\rm [keV]}\le 7-10$) surveys that absorption is negligible and 
the HXLF is calculated without taking absorption into account, simply by assuming a power-law spectra \citep[e.g.,][]{silverman08,yencho09,aird10}. 
For comparison, we also compute the case where all the AGNs have simple a $\Gamma=1.8$ power-law. The results of this case, with PDFs of 
the photometric redshifts are taken into account, are listed under ``$\Gamma=1.8$'' in Table \ref{tab:best} and the comparisons
with the ``Full'' case  are overplotted in Figs. \ref{fig:compare_model_z} and \ref{fig:compare_model_ll}.
As seen in these figures, the XLFs based on the simple spectral assumption of $\Gamma=1.8$ without absorption correction
is subject to underestimates of the XLF at low luminosities ($\log L_{\rm X}\lesssim 44$) at low redshifts ($z\lesssim 1$), by up 
to 50\%. If one uses a sample selected at $E\ga 5$ keV, however, this systematic error is expected to be reduced dramatically
(S. Fotopoulou et al. 2015, submitted to A\&A). 

{ In principle, the results depend on the $N_{\rm H}$ function and its evolution. Using the 14-195 keV selected {\sl Swift} BAT AGNs 
with detailed X-ray spectroscopy at $E<10$ keV and detailed modeling of selection effects, the $N_{\rm H}$ function derived by U14 is robust, 
especially at low redshifts, and in agreement with that of \citet{burlon11} derived using a larger sample of {\sl Swift} BAT AGNs. 
The fraction of absorbed AGNs among CTN AGNs is observed to increase with redshift \citep[e.g.][]{hasinger08}. However, it is not clear 
whether the increase of absorbed fraction saturates at $z\sim 2$ or keeps increasing to higher redshifts, and \citet{hasinger08} commented
that a model where the absorption fraction keeps increasing at $z>2$ is also marginally acceptable.  In our analysis above, we have 
used the U14 result, where the fraction of $22<\log N_{\rm H}<24$ AGNs among CTN AGNs at $\log L_{\rm X}=43.75$ grows as 
$\propto (1+z)^{0.45}$ at ($z<2.0$) and stays constant at higher redshift. For a sensitivity check, we assume that the absorbed fraction
grows as  $\propto (1+z)^{0.32}$ up to our sample limit, where the index has been chosen to have the same absorbed fraction at $z=3$. 
The results of the global fit under this assumption have not changed significantly, and all the parameters are consistent with our ``Full'' 
case within a fraction of $1\sigma$. This can be understood because, at high redshift, the 0.5-10 keV band, which is usually used for 
the spectral analysis or hardness ratio analysis to measure $N_{\rm H}$, corresponds to rest-frame $E>1.5$ keV at $z>2$. Thus it is already
not sensitive to measuring absorption. For the same reason, from the 2-10 keV flux, we measure rest frame higher energy luminosity, which 
has not been affected by absorption. Since we want to derive the intrinsic XLF, the measured X-ray luminosity is insensitive to the details
of the $N_{\rm H}$ function at high redshifts.}

\subsection{Detailed behavior of the XLF}

 Thanks to the addition of the COSMOS dataset, we have been able to trace the detailed behavior of the XLF
in the intermediate redshift/intermediate luminosity range. We describe a number of notable features.
\begin{itemize}
\item The low luminosity slope $\gamma_1$ flattens suddenly at $z\sim 0.6$. The slopes stay almost
  constant within each of the  $z<0.6$ and $z\geq 0.6$ regimes, having $\gamma_1\approx 1$ and $\approx 0.5$ 
  respectively. This sudden flattening of the slope practically excludes ILDE/LADE models, where the shape 
  of the XLF stays unchanged over redshift, as a global expression. The high redshift faint end slope is
  consistent with those of optically selected QSOs at $1\la z \la 3.5$ ($\gamma_1\sim 0.45$) measured by 
  \citet{bongiorno07} and $z\sim 4$ QSOs by \citet{ikeda11} ($\gamma_1=0.7\pm 0.2$). Note, however,  
  that the faint end slope measured by  \citet{gilkman10} is significantly steeper ($\gamma_1=1.3\pm 0.2$ 
  at $z\sim 4$).
\item The high luminosity slope stays almost constant ($\gamma_2\approx 2.7$), with the exception of
  two redshift shells at $0.6<z<0.8$ and $1.0<z<1.2$, which have flatter slopes $\gamma_2\approx 2.0$. 
  For the shell between these ($0.8<z<1.0$), $\gamma_2$ is consistent with those for the rest of the shells.
  We note that the high luminosity end of the XLF is poorly sampled at these redshifts and the apparently low 
  $\gamma_2$ values may be caused by the higher weights at luminosities just above the break and less weight 
  at highest luminosities. Thus there is no evidence for systematic change of the high luminosity
  slope with redshift. The upcoming {\sl eROSITA} mission will provide better sampling of this regime, 
  which will enable us to see if the high end slope is constant over all redshifts.
\item  The addition of the new data including COSMOS revealed new details about AGN downsizing. 
  The number density/emissivity growth curves have two distinctive breaks, one at $z_{\rm b1}\approx 1$ and the 
  other at $z_{\rm b2}\approx 2.7$, with a ``plateau'' between them. In any luminosity class, a rapid 
  rise of number density with cosmic time is observed at $z_{\rm b2}$ (except the lowest luminosity bin) and a rapid decline 
  is observed after $z_{\rm b1}$. At low luminosities, ($\log L_{\rm X}\la 44$),  the peak is at $z_{\rm b1}$ and high luminosities, 
  ($\log L_{\rm X}\la 44$),  $z_{\rm b2}$ becomes the peak. Also the position of the first break moves from lower 
  to higher redshifts with luminosity up from $z\sim 0.6$ to $z\sim 1.4$ up to $\log L_{\rm X}\la 44.5$. No evidence 
  for a change in  $z_{\rm b2}$ with luminosity is observed. 
\end{itemize} 
 
  Our results are in excellent agreement with those of U14 as seen in Fig. \ref{fig:compare_mdls}(a).
 Small excesses at the level of $\lesssim 0.2$ dex at a few data points can be well explained by cosmic variance. 
 Especially our ${\rm a\,few}\times 10^{15}\la S_{\rm X,obs}\la 10^{-14}$ ${\rm erg\,s^{-1}\,cm^{-2}}$ range is dominated by 
 the COSMOS data, which may be affected by large scale structures seen as redshift spikes 
 \citep{gilli09,brusa10,civano12}. Other possible differences at the same level may be explained in 
 the treatment of the probability density distribution of the photometric redshifts.
{ Our results at the highest redshift shell $z>3$ as well as the redshift evolution factor ($p_3$) are
consistent with the recent results by \citet{vito14}.}

\subsection{Comparison with Theoretical Models}

 There are a number of attempts to reproduce/understand the down-sizing behavior of AGN evolution within
cosmological simulations and semi-analytical models. In order to understand the implication of our results
to the physical scenarios on accretion processes by comparing observations and theory, we should 
consider which features of the theoretical predictions are a consequence of the assumed physical picture 
and which features are a consequence of adjusting free parameters to match the AGN luminosity functions
in various redshifts.

\begin{figure}
\begin{center}
\resizebox{!}{!}{
  \includegraphics[width=\hsize]{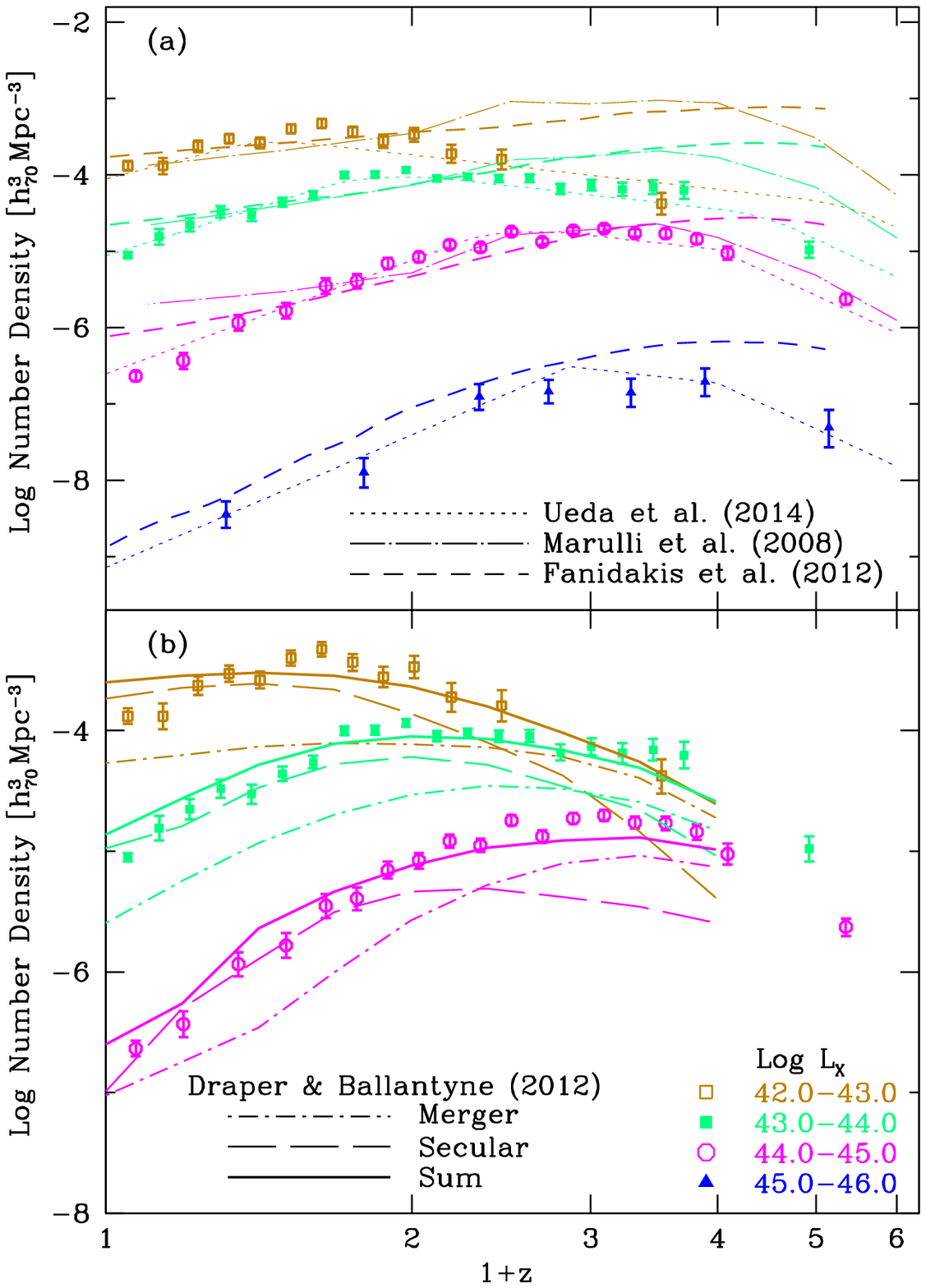}
}
\end{center}
\caption{(a) Our luminosity-class divided number density evolution curves (binned data points with errorbars
  from Fig. \ref{fig:ndens}a) are compared with the LDDE model from recent comparable work by \citet{ueda14} 
  (dotted lines). Also results from recent semi-analytcal models from literature from \citet{marulli08} (dot-dashed lines) 
  and \citet{fanidakis12} (dashed lines) are overplotted. Luminosity classes for the data points and model curves
  are indicated by colors as labeled. See text for detailed conversions of their published results to the 2-10 keV 
  density curves. Model by \citet{marulli08} does not reach the highest luminosity bin 
  and therefore we are not able to plot $\log L_{\rm X}=45-46$ bin. (b) The same number density curves are
  compared with the model by \citet{draper12} (solid lines) with separate contributions from merger (dotted lines)
  and secular processess (dashed lines). (Color versions of these figures are available in the electronic
  version.)\label{fig:compare_mdls}}
\end{figure}

 In their attempt to reproduce the faint end AGN LF, \citet{degraf10} considered a BH growth by infall
of surrounding gas via \citet{bondi52} accretion, which some authors call ``radio mode'' or
``hot halo mode'', and merging with other BHs in their simulations
including hydrodynamic gas component. They compared their results with luminosity-class divided number density evolution curves 
of soft \citep{hasinger05} and hard X-ray \citep{ueda03}. They recognize that their model overproduces number densities 
of X-ray selected AGNs at $z>1$. We compare our number density evolution with two semi-analytical models of  
\citet{marulli08} and \citet{fanidakis12}, where we have been able to convert their published results to the density curves 
in our 2-10 keV luminosity classes in  Fig.~\ref{fig:compare_mdls}(a).
This figure shows that the overproduction is also seen in these semi-analytical models (SAMs). 

 Since these authors did not provide the number density curves in the 2-10 keV range,
we explain the derivation of the luminosity-class divided 2-10 keV number density curves from their published 
figures. For \citet{marulli08}, we use their Fig. 7 and take the ``best'' bolometric luminosity 
function model. The bolometric LF has been converted to 2-10 keV XLF model using the bolometric 
correction by \citet{hopkins07} and integrated over our luminosity classes. For \citet{fanidakis12},
we use their soft (0.5-2 keV) X-ray ``before obscuration (total)'' curves provided in their Fig. 18 (b)
instead of their ``after obscuration (visible)'' to compare with our 2-10 keV density evolution curves. We 
convert their ``before obscuration'' to those for 2-10 keV assuming a photon index of $\Gamma=1.9$, 
which is representative of unobscured AGNs. Since they use slightly different cosmological parameters,
we also convert the density curves to our adopted cosmology to overplot in Fig.~\ref{fig:compare_mdls}(a).  

 In both cases, we see overpredictions of AGN number density at highest redshift at low luminosity bins.
As AGN model components, the \citet{marulli08} model took into account the ``radio'' 
mode accretion and merger driven ``quasar'' mode. The \citet{fanidakis12} model took into account the 
``radio'' and the ``starburst'' modes, where the ``starburst mode'' includes merger-driven ``quasar mode'' 
and secular (disk instability) components.  
\citet{marulli08} considered a number of AGN light curve models and searched for the one that fits 
well with the bolometric LF of \citet{hopkins07}, thus had freedom in adjusting to observations. 
On the other hand, \citet{fanidakis12} had four free parameters to fit to the LF. Except for the duty cycle,
which is adjusted to the global normalization of the LF, the parameters are physically motivated. 
The \citet{fanidakis12} model overpredicts the number densities at high redshift at 
all luminosity classes. In their comparison with the 0.5-2 keV XLF, they took into account the effects of luminosity 
dependence and redshift evolution of obscured/type 2 AGN fraction $f_2$ by \citet{hasinger08}. 
For the redshift evolution of $f_2$,  \citet{hasinger08} proposed a model with $f_2 \propto (1+z)^{0.62}$ up to $z=2.06$ 
and which stays constant at higher redshifts as preferred, while \citet{fanidakis12} used a marginally acceptable
model with $f_2 \propto (1+z)^{0.48}$ without saturation, and thus their obscured fraction keep increasing
to higher redshifts. The latter form of the obscured fraction evolution was needed in the model of 
\citet{fanidakis12} for satisfactory consistencies with the optical QSO and soft XLF to their model.
On the other hand, the high-z overprediction of the 2-10 keV number densities show that they indeed 
over predicted high redshift AGN population and over-corrected for the effects of obscuration to match
with the 0.5-2 keV number density/optical QSOs. The tendency of overprediction of high
redshift low luminosity/mass population is also a problem with predicting stellar mass function
of galaxies \citep[e.g.][]{guo11}, the reason for which is still unclear. On the other hand, if their basic 
prediction on high redshift accretion is correct, the discrepancy may be an indication of a large population 
of highly Compton-thick AGNs at high redshifts. { The high redshift over-prediction problem is
also present in the recent model by \citet{sijacki14} at all luminosities.}
  
 Another interesting model is by \citet{draper12}, where they considered merger and secular modes
of AGNs triggering. Their work was not directly based on cosmological simulations. Instead they 
took simple formulae for the evolution of galaxy number density with stellar mass above some threshold 
value, gas fraction, and merger rate from the literature. They then considered secular and merger
driven AGN triggering followed by a parameterized model of luminosity decay.
 Their model 2-10 keV number density evolutions are overplotted with our data in 
Fig.~\ref{fig:compare_mdls}(b), which has been derived from their Fig. 7. Since their Fig. 7 shows curves
for $\log L_{\rm X}>42$, $>43$, and $>44$, we draw their model curves of $\log L_{\rm X}=42-43$ and $43-44$ curves
by subtracting their $>43$ curve from $>42$ one and $>44$ curve from $>43$ one respectively. For the  
$44-45$ curve, we use their $>44$ curve, since the contribution of the $\log L_{\rm X}>45$ AGNs to the number 
density is only $\sim 0.2\%$ of $>44$ AGNs for $\gamma_2=2.7$. We see that their ``sum'' of the contributions of 
the two processes is in good agreement with our data. Since they have three free parameters to adjust 
in their AGN light curves, which are the timescale,  slope of the decay, and the peak Eddington ratio, they had much 
freedom in adjusting to observed XLFs. Thus it is no surprise that their theoretical curves agree well with our 
observation. However, they had difficulties in reproducing the hard X-ray number density curve with only one mode, 
i.e., merger only or secular only. Their summed curves show the flat-top (two break) structures, where
the slope in the $z_{\rm b1}<z<z_{\rm b2}$ range increases from negative to positive with luminosity.  

\section{Conclusions}
\label{sec:conc}

 We have investigated the detailed behavior of the intrinsic 2-10 keV X-ray luminosity function (XLF) of active
galactic nuclei (AGNs) from a combination of samples in various depths/width. We summarize our conclusions.

\begin{enumerate}
\item A total of $\sim 3200$ unique X-ray AGNs from a combination of wide to deep samples covering the redshifts of 
  $0.015<z<5.8$ and six orders of magnitude in flux have been used to investigate the behavior of the XLF. 
  In particular, we present here the XLF from X-ray sources from both
  of the {\sl XMM} and {\sl Chandra} COSMOS surveys, which comprise about 43\% of our global sample.
\item Recent $N_{\rm H}$ functions and realistic spectral templates are fully incorporated in the maximum-likelihood
  fitting to analytic expressions as well as the $N_{\rm obs}/N_{\rm mdl}$-estimated binned intrinsic XLF. The photometric 
  redshift probability density distributions (pz-PDFs) of the COSMOS and Lockman Hole samples as well as secondary 
  photometric redshifts of CDF-S are also considered.
\item We present a full parametric luminosity-dependent density evolution (LDDE) model, and redshift-shell and
  luminosity class separated models, as well as the binned XLF using the $N_{\rm obs}/N_{\rm mdl}$ estimator.
\item The low luminosity end slope of the XLF flattens suddenly at $z>0.6$. This behavior practically excludes
 any global expression of XLF evolution that assumes an unchanged XLF shape over cosmic time.
\item We investigate the net effects of ignoring probability density distribution (pz-PDF) of photometric redshifts. 
  {The effect of including pz-PDF into the analysis does not alter the final XLF results significantly.} 
\item We investigate the effects of assuming a simple AGN spectrum of $\Gamma=1.8$. Under this simple assumption, the
  XLF is subject to underestimates by up to $\sim 50$\% in the low redshift low luminosity regime.
\item The detailed behaviors of AGN down-sizing have been revealed. A clear two-break structures has
  been revealed in the number density evolution curves at every luminosity class above $\log L_{\rm X}>43$.
  This behavior is qualitatively consistent with a two-mode AGN evolution involving major merger and secular 
  processes.
\item Most current semi-analytical models of AGNs over-produce AGN number densities in the high
  redshift, low luminosity regime. Their semi-analytical treatment of accretion processes may have to be revised
  to reproduce our results. Alternatively, if their accretion scenario is correct, this may suggest a population
  of heavily Compton-thick AGNs in this regime. 
\end{enumerate}

\begin{acknowledgements}
 This work is supported by UNAM-DGAPA Grant PAPIIT IN104113 and  CONACyT Grant Cient\'ifica B\'asica \#179662 (TM).
Also acknowledged are the FP7 Career Integration Grant ``eEASy'' CIG 321913 (MB),
financial support from INAF under the contracts PRIN-INAF-2011 and PRIN-INAF-2012 (NC,AC,RG), and  
the Grant-in-Aid for Scientific Research 26400228 (YU) from the Ministry of Education, Culture, Sports, Science 
and Technology of Japan (MEXT). We gratefully acknowledge the contributions of the entire COSMOS collaboration consisting 
of more than 100 scientists. More information about the COSMOS survey is available 
at http://www.astro.caltech.edu/~cosmos. The authors thank Len Cowie for allowing us to use their unpublished 
spectroscopic redshifts and Murray Brightman for his advice on CTK AGNs.
\end{acknowledgements}

Facilities: \facility{XMM},\facility{CXO},\facility{ASCA},\facility{Swift}

\end{document}